
\lineskip=6pt minus 2pt
\lineskiplimit=3pt
\magnification=1200

\def\simle{{}^<_{\sim}}
\def\simge{{}^>_{\sim}}
\def\oh{\Omega_b h^2}
\def\h2{(^2{\rm H/H})}
\def\he3{(^3 {\rm He/H})}
\def\li7{(^7 {\rm Li/H})}
\def\dc{\Delta_{cr}}
\def\ooo{\Omega_b/\Omega_{diff}}

\centerline{\bf NUCLEOSYNTHESIS IN THE PRESENCE OF}
\vskip 0.08in
\centerline{\bf PRIMORDIAL ISOCURVATURE BARYON FLUCTUATIONS}
\vskip 0.4in
\baselineskip=14pt plus 2pt
\centerline{{\sl Karsten Jedamzik$^1$} and {\sl George M.
Fuller$^2$}}
\vskip 0.2in
\centerline{$^1$Physics Research Program}
\centerline{Institute for Geophysics and Planetary Physics}
\centerline{University of California}
\centerline{Lawrence Livermore National Laboratory}
\centerline{Livermore, CA 94550}
\vskip 0.15in
\centerline{$^2$Department of Physics}
\centerline{University of California, San Diego}
\centerline{La Jolla, CA 92093-0319}
\vskip 0.4in
\centerline{\bf Abstract}
\vskip 0.06in
We study big bang nucleosynthesis in the presence of large
mass-scale,
non-linear entropy fluctuations. Overdense regions, with masses above
the local
baryon-Jeans mass, are expected to collapse and form condensed
objects. Surviving
nucleosynthesis products therefore tend to originate from underdense
regions.
We compute expected surviving light element ($^2$H, $^3$He, $^4$He,
$^7$Li)
abundance yields for a variety of stochastic fluctuation spectra. In
general,
we find that spectra with significant power in fluctuations on length
scales below that of the local baryon Jeans mass produce
nucleosynthesis yields which are in conflict with observationally
inferred primordial abundances. However, when this small scale
structure is absent or suppressed, and the collapse efficiency of
overdense
regions is high, there exists a range of fluctuation spectral
characteristics
which meet all primordial abundance constraints. In such models
abundance
constraints can be met even when the pre-collapse baryonic fraction
of the
closure density is $\Omega_b\approx 0.2h^{-2}$ ($h$ is the Hubble
parameter
in units of 100 km\ s$^{-1}$Mpc$^{-1}$). Nucleosynthesis in these
models
is characterized by high $^2$H/H and low $^4$He mass fraction
relative to a
homogeneous big bang at a given value of $\Omega_bh^2$. A potentially
observable
signature of these models is the production of intrinsic primordial
abundance variations on baryon mass-scales up to
$10^{10}M_{\odot}-10^{12}
M_{\odot}$.
\vskip 0.3in
\centerline{\it Subject headings: cosmology: theory - early universe
-}
\centerline{\it theory - nucleosynthesis, abundances - theory - large
scale
structure: general}
\vfill\eject

\baselineskip=16pt plus 2pt
\centerline{\bf 1. Introduction}

In this paper we calculate the nucleosynthesis to be expected in the
presence of primordial isocurvature baryon number (hereafter, PIB)
fluctuations.
Such fluctuations recently have been proposed as possible seeds for
large
scale structure formation (Peebles 1987ab; Suginohara \& Suto 1992;
Cen, Ostriker, \& Peebles 1993). Density fluctuations in PIB models
are
essentially {\it entropy} fluctuations. In an earlier series of
papers
(Jedamzik \& Fuller 1994; Jedamzik, Fuller, \& Mathews 1994) we have
examined
the evolution and nucleosynthesis effects of small mass-scale
nonlinear entropy
fluctuations. In the present paper we extend our nucleosynthesis
study to larger
mass-scale fluctuations, including those relevant for PIB models.

Proposed PIB models invoke a spectrum of entropy fluctuations
characterized
by increasing amplitude with decreasing fluctuation mass scale. In
these models
the relationship between the fluctuation amplitude, $\delta\rho
/\rho$ , and the
mass contained inside the region of the fluctuation, $M$ , can be
written as
$${\delta\rho\over\rho}\sim M^{-{1\over 2}-{n\over 6}}\ ,\eqno(1)$$
where $n$ is a spectral index. Spectra of this form are fully
specified
by giving the index, $n$ , and the mass scale, $M_{unity}$ , on which
$\delta\rho /\rho =1$. Typical spectral indices employed in PIB
models
include $n=0$ or $n=-1$, where accompanying values of $M_{unity}$ are
$$10^{11}\ {^>_{\sim}}\ {M_{unity}\over M_{\odot}}\ {^>_{\sim}}\
10^6\ .\eqno(2)$$

We have no compelling microscopic theory for how PIB fluctuations
could be
generated in general, though there are some cogent suggestions.
Yokoyama
\& Suto (1991) and Dolgov \& Silk (1992) have proposed microscopic
mechanism for the generation of baryon number (entropy) fluctuations.
These
schemes exploit quantum fluctuations in a CP-violating angle to
effect a spatial
variation in baryon-to-photon number. The resulting fluctuations can
be on the
large mass-scales of interest in PIB models if the baryon-generating
process
occurs before, or during, an inflationary epoch.

A common, though not necessarily inevitable, feature of inflationary
baryogenesis models is the production of both matter and antimatter
domains
separated by astrophysically significant length scales. The
production of
antimatter domains can be avoided if the baryon number fluctuation
amplitudes
are below unity on all length scales. In this case, however, the
resulting light
element abundance yields from nucleosynthesis will be only slightly
altered
from abundance yields in homogeneous big bang nucleosynthesis at the
same
average baryon-to-photon ratio $\eta$ (for a discussion of small
amplitude
fluctuations in $\eta$ and primordial nucleosynthesis yields see
Epstein \& Petrosian 1973).

Though large scale structure considerations in PIB models utilize
only the
large mass-scale linear ($\delta\rho /\rho <1$) end of the spectrum
in
equation (1), successful models may well require some small
mass-scale,
nonlinear ($\delta\rho /\rho > 1$) structure in order to produce
collisionless
dark matter and/or to effect late re-ionization of the universe
(Cen {\it et al.} 1993). Of course, the biggest hurdle which PIB
models
face may well be stringent observational limits on cosmic background
radiation anisotropy (Gorski \& Silk 1989; Chiba, Sugiyama, \& Suto
1993;
Hu \& Sugiyama 1994).

If any PIB model for structure formation must invoke a nonlinear
lower
mass-scale fluctuation tail in a spectrum like that in equation (1),
or
if such a tail is an inevitable consequence of some microscopic
fluctuation
generation mechanism, then primordial nucleosynthesis effects may
well
provide constraints on the model. These constraints could be
complimentary
to cosmic-background-anisotropy constraints, as nucleosynthesis
probes a
different end of the fluctuation spectrum. In this paper we aim to
see
under what circumstances such constraints on PIB-like models could be
found.
Our study and our derived constraints will generally apply to {\it
any} models
which have nonlinear entropy fluctuations - not just PIB-like models
for large
scale structure formation.

Studies of homogeneous big bang nucleosynthesis (HBBN) only allow a
narrow
range for the fraction of the closure density that can be contributed
by baryons
$$ 0.01\ {^<_{\sim}}\ \Omega_bh^2\ {^<_{\sim}}\ 0.015\ ,\eqno(3)$$
where $h$ is the Hubble parameter in units of 100
kms$^{-1}$Mpc$^{-1}$
(cf. Walker {\it et al.} 1991; Smith, Kawano, \& Malaney 1993).
Models of
inhomogeneous big bang nucleosynthesis (IBBN) with sub-horizon scale
nonlinear entropy fluctuations cannot significantly change the
conlusions of
HBBN as to the upper limit on $\Omega_bh^2$ (cf. Jedamzik, Fuller, \&
Mathews
1994; Thomas {\it et al.} 1994).

However, there has been a longstanding suggestion that large mass
scale
entropy fluctuations might provide a way to circumvent the HBBN upper
bound on $\Omega_bh^2$ (Harrison 1968; Zeldovich 1975; Rees 1984;
Sale \&
Mathews 1986). In these schemes regions of relatively large
baryon-to-photon
ratio ($\eta$) collapse to form inert remnants and therefore remove
their
\lq\lq bad\rq\rq\ nucleosynthesis products from the primordial
medium.
The nucleosynthesis products which {\it survive} in these models
reflect
the relatively lower baryon density of regions which do not collapse.
In principle, the pre-collapse $\Omega_bh^2$ could be considerably
larger
than the value of this quantity (presumbly $\Omega_bh^2\approx
0.01-0.015$)
which characterizes the low density medium which surrounds collapsed
regions.

Indeed, studies of the fate of large mass entropy fluctuations in the
epoch
between primordial nucleosynthesis and some time after recombination
indicate
that overdense regions would collapse with high efficiency
(Hogan 1978, 1993; Kashlinsky \& Rees 1983; Loeb 1993). These studies
show
that the evolution of large entropy fluctuations is dominated by
photon-electron
Thomson drag (Hogan 1993; Loeb 1993). Any fluctuation with a baryon
mass larger
than about
$$M_J^b\approx 3\times 10^5M_{\odot}\biggl({\Omega_bh^2\over
0.1}\biggr)^{-{1\over 2}}\ ,
\eqno(4)$$
will shrink slowly, with Thomson drag providing an efficient
mechanism for
cooling the collapsing protons and electrons, and also damping
rotation. The end
result of this process will be the production of a condensed object,
either
a black hole or possibly, if fragmentation occurs, many small brown
dwarfs
(Kashlinsky \& Rees 1983; Hogan 1993). Note that the mass scale in
equation (4)
is essentially coincident with the {\it local baryon Jeans mass}. Of
course,
the true Jeans mass in a radiation dominated environment is close to
the
horizon mass. By local baryon Jeans mass we mean the effective Jeans
mass
one would calculate inside a fluctuation by neglecting all photon
stresses.

We note, however, that the studies of Harrison (1968), Zeldovich
(1975),
Rees (1984), and Sale \& Mathews (1986) provide what are at best
simplistic
treatments in their calculations of nucleosynthesis yields in the
presence
of collapsing regions. In fact, all of these studies except Sale \&
Mathews
(1986) treat the distribution of $\eta$ in a two-phase, bi-modal
fashion.
All of these studies neglect the significant complication that
fluctuations
may reside inside larger mass-scale fluctuations, the so called
cloud-in-cloud
problem. None of these prior studies addresses the nucleosynthesis
effects of
fluctuations below that in equation (4) - such fluctuations are
damped by
expansion against Thomson drag (cf. Hogan 1993; Jedamzik \& Fuller
1994), but
on time scales that allow significant nucleosynthesis effects
(Alcock {\it et al.} 1990; Jedamzik, Fuller, \& Mathews 1994).

Recently, Gnedin, Ostriker, \& Rees (1994) have re-examined the
problem of
nucleosynthesis with collapsing entropy fluctuations. These authors
provide
a more sophisticated numerical treatment of nucleosynthesis from
regions of
varying density and, additionally, attempt to take account of light
element
\lq\lq reprocessing\rq\rq\ effected by accretion on black holes
(Gnedin \& Ostriker 1992). They conclude that significant power on
fluctuation
scales below the limit in equation (4) could lead to unacceptable
nucleosynthesis which, in turn, could wreck any attempt to employ
collapsing
high-mass-scale fluctuations to circumvent HBBN bouns on $\Omega_b
h^2$.
They do, however provide for several loopholes in this conclusion:
phase
correlations in fluctuations; and invocation of finely tuned IBBN
scenarios
for small mass-scale fluctuations.

However, Gnedin, Ostriker, \& Rees (1994) do not attempt to take
account of the
effects of the cloud-in-cloud problem. Furthermore, their suggestion
that
baryon diffusive effects on small scales may provide a loophole on
the
$\Omega_bh^2$ limit is suspect, as it requires an extreme, though not
impossible, degree of fine tuning.

In this paper we attempt a detailed numerical treatment of
nucleosynthesis with
various stochastic entropy fluctuation spectra. We explicitly treat
the
cloud-in-cloud problem. We find that any significant power in
stochastic
fluctuations on mass-scales smaller than that in equation (4)
inevitably
leads to overproduction of $^4$He and $^7$Li relative to
observationally
inferred limits on these abundances. Furthermore, we argue that
persistence
of fluctuations down to mass scales on which baryon diffusive effects
are
significant is unlikely to change our conclusions. Finally, we show
that
only under a restrictive set of circumstances (i.e. no significant
power
in fluctuations on mass scales below $M_J^b$ and high collapse
efficiency for
scales above $M_J^b$) it is possible to evade HBBN bounds on
$\Omega_bh^2$.
We note that evasion of the bound on $\Omega_bh^2$ in these models
also usually
requires that the primordial abundance of $^7$Li be close to the
Population I vaule
of $^7$Li$/$H $\approx 10^{-9}$.

In Section 2 we discuss our model for fluctuation evolution and
nucleosynthesis and the numerical techniques employed. In Section 3
we present
results for nucleosynthesis yields for various fluctuation spectra
and
assumed fluctuation evolution parameters. We compare these to
observationally
inferred abundances. We give conclusions in Section 4.

\vfill\eject
{\bf 2. Simulations of baryon-to-photon fluctuations}
\vskip 0.1in

In this section we discuss our numerical calculations of fluctuation
evolution
and nucleosynthesis. The basis of our models is a stochastic
distribution of
fluctuations in baryon-to-photon ratio on various scales. These
scales will be
selected in such a manner as to roughly approximate the density
distribution
in an overall spectrum like that in equation (1). Note, however, that
much is
hidden in a simplistic power law density distribution like equation
(1).
Going from equation (1) to a numerical representation of fluctuation
amplitude
and mass-scale distribution is not a unique procedure. In what
follows we
present our numerical approach to this problem and discuss the
statistics of
our numerically generated distributions of baryon-to-photon number.

\vskip 0.15in
{\bf 2.1. The numerical models}
\vskip 0.1in

As a first step, we generate a stochastic distribution of
baryon-to-photon number by employing a spatially inhomogeneous
gaussian random variable, $A(x)$.
This gaussian random variable can be Fourier decomposed to give,
$$A(x)=2\sum_k A_k {\rm cos}(kx+\phi_k)\ ,\eqno(5)$$
where $x$ is a spatial coordinate. In this expression the amplitudes
$A_k$ and phases $\phi_k$ are chosen randomly according to the
distribution
functions,
$$P_{\phi}(\phi_k)={1\over 2\pi}\quad \phi_k\in [0,2\pi ]\
,\eqno(6a)$$
$$P_{A}(A_k)={1\over\sqrt{2\pi}}{1\over\sigma_k}{\rm exp}\biggl(
-{1\over 2}{A_k^2\over\sigma_k^2}\biggr)\quad A_k\in [-\infty ,\infty
]\ .
\eqno(6b)$$
Note that our adoption of the distribution function in equation (6a)
implies
that there will be no correlations between the phases of different
Fourier modes - of course, this is the definition of a gaussian
random
variable. We adopt a power law dependence on wavenumber for the
variance,
$\sigma_k$,
$$\sigma^2_{k}\sim k^n\ ,\eqno(7)$$
where $n$ is a spectral index, not to be confused with that in
equation (1).

The spatially dependent baryon-to-photon ratio , $\eta (x)$ , is
taken to be a
function of the spatially fluctuating random variable $A(x)$.
We consider three different models:
$$\eta (x)=\eta_N A^2(x)\ ;\eqno(8a)$$
$$\eta (x)=\eta_N A^{10}(x)\ ;\eqno(8b)$$
$$\eta (x)=\eta_N \bigl( A^{10}(x)+a\bigr)\ ;\eqno(8c)$$
where $\eta_N$ and $a$ are constants. Note that the functions in
equations (8abc) were deliberately chosen to be positive-definite.
This choice guarantees that antimatter-domains are avoided.

Our choice of the spatial distributions in equations (8abc) is
not based on specific baryogenesis
scenarios. Rather, we have chosen these models simply to generate a
wide
variety of stochastic baryon number distributions. It remains to be
shown if
baryon number distributions of these, or similiar characteristics,
could
arise naturally during the evolution of the very early universe.

In this paper we analyze a one-dimensional analogue to a
three-dimensional distribution of entropy (or, equivalently, $\eta$).
By using a
one-dimensional distribution of $\eta$ we are able to
investigate a much wider range of mass scales in a single
simulation. The analogy to the three-dimensional theory is attained
by replacing the spectral index $n$ in equation (7) with $n/3$.
For example, $n=-3$ corresponds to a scale-invariant
Harrison-Zeldovich
spectrum in the three-dimensional theory; whereas $n=-1$ corresponds
to
the Harrison-Zeldovich spectrum in one dimension.

In what follows we
will always refer to the three-dimensional spectral index when we
wish to
characterize a particular one-dimensional distribution in $\eta$.
In a one-dimensional calculation we compute the \lq\lq mass\rq\rq\ of
a region
by multiplying \lq\lq density\rq\rq\ (or amplitude $\delta\rho
/\rho$) by the
length scale of the region. In three dimensions, masses are the
product of
$\delta\rho /\rho$ and a volume. The transformation of spectral
indices
$n\mapsto n/3$, insures that the functional dependence of $\delta\rho
/\rho$
on mass-scale (e.g. equation 1) that we derive in our one-dimensional
calculation is the same as that in an equivalent ($n/3 \mapsto n$)
three-dimensional case.

It is well known that diffusive and hydrodynamic processes occuring
during the epoch of primordial nucleosynthesis can significantly
alter the light-element abundance yields. Diffusive processes will
play an important role during nucleosynthesis when significant
small-scale fluctuations on mass scales between $M\sim
10^{-21}M_{\odot}$ and $M\sim 10^{-11}M_{\odot}$ are present.
Nuclear abundance yields in such scenarios will certainly be highly
geometry dependent so that a full three-dimensional treatment would
be needed. (Mathews {\it et al.} 1990; Meyer {\it et al.} 1991).
However, we will suggest below that, for a truly
stochastic baryon number distribution (e.g. minimal phase
correlations between different Fourrier modes of the distribution),
the presence of such small-scale fluctuations cannot be reconciled
with the observationally inferred primordial abundance limits. We
will therefore not be concerned with diffusive processes.

In the case where diffusive and hydrodynamic processes during
nucleosynthesis are unimportant, average light-element abundance
yields are determined by a weighted average over the standard
homogeneous big bang yields of separate regions at different
baryon-to-photon ratios. In this study we obtain the light-element
abundance yields by employing the homogeneous big bang
nucleosynthesis
code of Wagoner, Fowler, \& Hoyle (1967) as updated by Kawano (1992).

In our calculations the light-element contributions from overdense
regions,
which are expected to collapse and form compact objects,
are excluded from the abundance average.
We choose all regions which are overdense on average by some
critical amount $\Delta_{cr}$ to be \lq\lq destined\rq\rq\ for
collapse.
We define the overdensity parameter, $\Delta_{\lambda}$ , to be,
$$\Delta_{\lambda}=
{{\eta_{\lambda}(x)-{<\eta >}}\over{<\eta >}}\ .\eqno(9)$$
In the above expression
$<\eta >$ denotes the cosmic average baryon-to-photon ratio and
$\eta_{\lambda}(x)$ represents the average baryon-to-photon ratio
within a region of size $\lambda$ around space coordinate $x$.
In our numerical prescription only regions with $\Delta_{\lambda}=
\Delta_{cr}$ are taken to collapse. If a region is found to have
$\Delta_{\lambda}>\Delta_{cr}$
we look for a larger region, which contains the original, for which
$\Delta_{\lambda}=\Delta_{cr}$. This larger region is taken to
collapse.
In other words, we take the {\it largest} scale for which
$\Delta_{\lambda}=\Delta_{cr}$ to collapse and take everything inside
to
oblivion.

The above prescription for determining which regions collapse
is a widely used analytic tool in the study
of structure formation (cf. Press \& Schechter 1974).
This procedure is also confirmed by comparison of the
analytic results to numerical simulations
(Efstathiou {\it et al.} 1988).
Note that the Press-Schechter analysis has been recently modified
in order to account properly for the cloud-in-cloud problem (Jedamzik
1994).

The epoch at which
an overdense region collapses is determined, for a given cosmological
model,
predominantly by its initial average overdensity.
For example, the formation of supermassive black holes in PIB models
requires early
collapse and high initial overdensities (Hogan 1993).

Partly, this requirement arises because
Thomson-drag on background photons becomes a less efficient mechanism
for entropy transport and rotational damping at lower cosmic
temperature.
A certain determination of the fate of a collapsing cloud would
require a full
three-dimensional hydrodynamic simulation. Determining the fate
of overdense regions is important,
since we need to know whether or not to count their freeze-out big
bang
nucleosynthesis products in the final abundance yields of the
diffuse,
\lq\lq uncollapsed\rq\rq\ , surviving background of primordial gas.

The value of $\Delta_{cr}$ where the efficiency of collapse
becomes large depends on the cosmological
model and the ionization history of the universe (Hogan 1993).
We define the collapse efficiency parameter, $f$ , to be the fraction
of regions
with $\Delta_{\lambda}=\Delta_{cr}$ which actually do collapse.
In this
paper we will simply treat $\Delta_{cr}$ and $f$ as parameters to be
varied. We will determine the sensitivity of our calculated
light-element abundance yields, and ratio of diffuse baryons
to dark baryons, to these parameters.

To determine the light-element abundances
in the presence of fluctuations it is not sufficient to
consider only the average densities of regions. Rather, in principle,
a detailed
knowledge of the baryon number distribution on all scales is
required.
Toward meeting this requirement, we define $P_{\eta}(x)$ to be the
probability
that the region at point $x$ has baryon-to-photon ratio $\eta$. We
will find,
ultimately, that all small mass scale fluctuations $(M<M_J^b)$ would
have to be
suppressed in order that a PIB-like fluctuation spectrum would be
able to
produce nucleosynthesis consistent with observational constraints. If
we
utilize this result, and simply assume that on small enough scales
fluctuation
amplitudes go to zero, then we can exploit the overall translational
symmetry of
the universe to write $P_{\eta}(x)=P(\eta )$. In other words the
probability
to find a region with baryon-to-photon ratio $\eta$ is independent of
$x$;
mathematically $P(\eta )$ is the probability of finding any point to
have
baryon-to-photon ratio $\eta$. By translational symmetry in this
argument
we mean that any one {\it large} region of the universe (large enough
to have
$\bar\eta$) must be fully equivalent and similar to any other such
region.

Note that writing $P(\eta )$ makes sense only if there is a small
scale cutoff
in structure. In our numerical calculations any scale smaller than
this
cutoff scale is taken to be homogeneous. The probability for finding
any
such spatial zone to have baryon-to-photon ratio $\eta$ is $P(\eta
)$. The
probability that a given scale $\lambda$ has average baryon-to-photon
ratio
$\eta$ is defined to be $P_{\lambda}(\eta )$. Clearly, with this
definition,
$P_{\lambda}(\eta )=P(\eta )$ for all $\lambda \leq \lambda_c$, where
$\lambda_c$ is the cutoff scale. We will discuss the statistical
relationship between $P_{\lambda}(\eta )$ and $P(\eta )$ for larger
scales
below.

It is quite important for what follows to note that $P(\eta )$, and
any
mass-based statistical criteria for describing mass distributions,
will be
invariant under the transformation from one dimension to three
dimensions as
outlined above. An example of an invariant mass-based distribution
function
is the following: the distribution of masses which have overdensities
equal to $\Delta_{cr}$.

We wish to stress the necessity of a numerical treatment
for the reliable determination of nuclear abundance yields.
Furthermore, there are many subtle pitfalls involved in estimating
abundance
yields in an inhomogeneous environment.
For example, it is not
adequate to simply average over nucleosynthesis yields resulting from
all
regions below some threshold, $\eta\leq\eta_{cr}\ $.
This procedure is inadequate since such
regions could possibly be underdense clouds within collapsing
overdense regions. As another example,
an analytic computation of $P_{\lambda}(\eta)$ is difficult
(impossible) since
the transformation in equations (8abc) introduces phase correlations
between
the different Fourier modes of the baryon number distribution. We
therefore
expect the stochastic baryon number distribution of our model to
have a non-gaussian character.

We have simulated three types of stochastic baryon number
distributions by applying the transformation in equation (8abc) to a
gaussian
random variable. For these simulations we generated $10^5$ Fourier
modes of the gaussian random variable. The wavevectors of these
Fourier modes were determined by applying periodic boundary
conditions in a box extending from zero to one. A small part of
these distributions are shown in Figures 1abc. Note that the value
of $10^0$ on the logarithmic abscissa corresponds to the cosmic
average
baryon-to-photon ratio, $<\eta >$ , in all three figures.

In Figure 1a most of the peaks in the distribution of $\eta$ are seen
to be
between
values of 5 and 10 (e.g., five to ten times average density);
whereas, the distribution shown in Figure 1b exhibits peaks with
values between 10 and 100. The distribution of $\eta$ in
Figure 1c also
exhibits large overdense peaks. However, in this distribution there
are no
regions with very low density. The dotted boxes in these figures
indicate the regions which are overdense by the critical amount,
$\Delta_{cr}$. These regions will be expected to collapse
very early on. It is obvious from the figures that such overdense
regions
often
are centered around very prominent peaks.
Note, however, that these overdense regions can include underdense
material at
small
baryon-to-photon ratios as well. We will analyze the nucleosynthesis
resulting from these baryon-to-photon number distributions in Section
3.

\vskip 0.15in
{\bf 2.2. Statistics of the baryon-to-photon number distributions}
\vskip 0.1in

In this section we will investigate the statistics of the
baryon-to-photon number
distributions generated by the prescription outlined in section 2.1.
Consider first a gaussian random variable $A$. The probability
distribution ${\bar P}(A)$ for finding a small region to have value
$A$ is given by the gaussian distribution,
$${\bar P}(A)={1\over\sqrt{2\pi}}{1\over\sigma_{tot}}{\rm exp}\biggl(
-{1\over 2}{A^2\over\sigma^2_{tot}}\biggr)\ ,\eqno(10a)$$
where
$$\sigma^2_{tot}={1\over 2}\biggl({L\over 2\pi}\biggr)
\int_0^{k_c}dk\ \sigma_k^2\ .\eqno(10b)$$
Note the relationship between these expressions and those for
$P_A(A_k)$ and $\sigma_k^2$ in equations (6b) and (7).
Here $(L/2\pi)^{-1}$ is the volume in wavevector space in which
there is one Fourier mode. Equation (10) makes use of the fact that
the
sum, $A$, of several normally distributed quantities (e.g., the
amplitudes of the uncorrelated Fourier modes, $A_k$) itself follows
a gaussian distribution. The square of the variance,
$\sigma_{tot}^2$ , of this gaussian
distribution in $A$ is then given by the sum of the squares of the
individual variances of the Fourier modes, $\sigma_k^2$.
Note that in equation (10b) we introduce a cutoff-wavevector $k_c$,
since
we assume that there are no small-scale fluctuations. In particular,
we take
$\sigma_k\approx 0$ for $k>k_c=2\pi /\lambda_c$.

With the help of equation (10a) we can
derive the probability function $P(\eta )$. Assuming a transformation
of the form $\eta (x)=\eta_N(A^{2m}(x)
+a)$ we obtain,
$$P(\eta )={1\over\sqrt{2\pi}}{1\over m\sigma_{tot}\eta_N}\biggl(
{\eta\over\eta_N}-a\biggr)^{{1\over 2m}-1}{\rm exp}\biggl\{{
-{1\over 2}\Bigl({\eta\over\eta_N}-a\Bigr)^{1\over
m}\over\sigma^2_{tot}}\biggr\}\ ,\eqno(11)$$
where $\eta$ falls in the range $\eta\in [a\eta_N,\infty]$.
In this expression note that $m$ is an integer index, while $\eta_N$
and $a$
are constants.

It
should be stressed that it is not possible to obtain the probability
function $P_{\lambda}(\eta )$
by simply employing equation (11) with a modified variance
$\sigma_{tot}\mapsto
\sigma\sim\int_0^{k}dk^{\prime}\sigma^2_{k^{\prime}}\ .$
This is because with a positive definite transformation such as that
in
equation (8abc) the average $\eta$ of a region is determined by the
larger-scale Fourier modes with $k^{\prime}<k$ as well as by the
smaller-scale
Fourier modes $k^{\prime}>k$. We therefore must resort to numerical
techniques in order to analyze the baryon-to-photon number
distribution on
larger scales.

We define the integrated variance in the distribution of $\eta$ on
mass scales
below the cutoff scale to be $\sigma (M<M_c)$. If $M_c$ is the mass
scale
corresponding to the cutoff wavevector $k_c$, then we can define
$$\sigma_{\eta}(M<M_c)=\biggl\{{1\over{<\eta
>}}\int_{a\eta_N}^{\infty}
d\eta \bigl(\eta -<\eta >\bigr)^2P(\eta )\biggr\}^{1\over 2}\
.\eqno(12)$$
The quantity
$\sigma_{\eta}$ determines the likely magnitude of fluctuation
amplitudes. For the transformation $\eta (x)=\eta_N\bigl(A^{2m}(x)+a
\bigr)$ we can derive, for example,
$$\sigma_{\eta}(M<M_c)={1\over {<\eta >}}{\eta_N\over\sqrt\pi}\bigl(
2\sigma^2_{tot}\bigr)^m\biggl[\sqrt\pi \Gamma (2m+{1\over 2})-
\Gamma^2(m+{1\over 2})\biggr]^{1\over 2},\eqno(13)$$
where the average baryon-to-photon ratio is given by
$$<\eta >={\eta_N\over\sqrt\pi}\biggl[\bigl(2\sigma^2_{tot}\bigr)^m
\Gamma (m+{1\over 2}) + a\Gamma ({1\over 2})\biggr]\ .\eqno(14)$$
If we apply equation (13) to the transformation in equation (8a) we
obtain
$\sigma_{\eta }(M<M_c)=\sqrt{2}$. By contrast, the fluctuation
amplitudes for the
transformations in equations (8bc) are larger and yield
$\sigma_{\eta}(M<M_c)\approx
27$. It should be noted that $\sigma_{\eta}$ is independent of
$\sigma_{tot}$.

We have investigated numerically the statistics of the
baryon-to-photon number
distributions on large scales. Figures 2abc show the distribution
functions $P_{\lambda}(\eta )$ on
different mass scales $M$ (or, equivalently, spatial scales
$\lambda$)
for the three transformations in equations (8abc).
In these figures the value of 1 on the abscissa corresponds to the
average baryon-to-photon ratio; whereas, the scale of the ordinate is
chosen in each case so that
the complete distribution of $P_{\lambda}(\eta )$ is displayed.
The lines on these figures are for the
distribution function $P_{\lambda}(\eta )$ on mass scales $M=M_c$
(solid), $M=10M_c$ (dotted), $M=100M_c$ (short-dashed), and
$M=1000M_c$ (long-dashed). We will determine the numerical value of
$M_c$ in the following section. The dashed-dotted line shows the
result from equation (11) for comparison. It can be seen
from these figures that the distribution
functions $P_{\lambda}(\eta )$ are highly non-gaussian on small
scales, but approach a gaussian distribution character on large
scales.
Note that the variance, $\sigma_{\eta}$ , decreases
as the mass scale increases.

A non-gaussian distribution $P_{\lambda}(\eta )$ on small scales
could
only result if there were phase correlations between the
different Fourier modes of the baryon-to-photon number distribution.
We note that phase
correlations are introduced in our procedure as a result of our
application of the transformation equation (8abc)
to the uncorrelated gaussian random variable. Similiar results
(e.g., non-gaussian distributions on small scales and gaussian
distributions on large scales)
have been obtained by Yamamoto {\it et al.} (1992). These authors
analyzed the
baryon-to-photon number distribution resulting from the inflationary
baryogenesis scenario proposed by Yokoyama \& Suto (1999).
That baryogenesis scenario assumed that the production of baryon
number
was proportional to the a trigonometric
function of a gaussian random variable, e.g.
$\eta (x)=\eta_0sin(A(x))$.

In Figures 3abc we show the fluctuation variance
$\sigma_{\eta}(M)$ as a function of mass scale for the three
different transformations in equations (8abc).
In each of these figures the dotted line gives $\sigma_{\eta}(M)$ in
the case
where the variance of the Fourier amplitudes is taken to be constant,
$\sigma_k^2=constant$. The dashed line in these figures corresponds
to
choosing $\sigma_k^2\sim k^{-2.4}$.
Here, the index $-2.4$ is the
\lq\lq three-dimensional\rq\rq\ index. For purposes of comparison in
these
figures we show the root-mean-square mass fluctuation expected for
a gaussian random variable,
$$\sigma_{\eta}(M)=\biggl({\delta M\over M}\biggr)_{\rm r.m.s.}=
\sigma_{\eta}(M_c)\biggl({M\over M_c}\biggr)^{-{1\over 2}-{n\over
6}}\ .\eqno(15)$$
The solid lines in these figures correspond to $\sigma_{\eta}(M)$
from
equation (15) with $n=0$ (lower line) and $n=-1.5$ (upper line).

\vskip 0.15in
{\bf 2.3. Normalization of the baryon-to-photon number distributions}
\vskip 0.1in

We can now determine the absolute magnitude of
the cutoff mass-scale, $M_c$ , for models with different spectral
indices , $n$ , and variances, $\sigma_{\eta}(M_c)$. Primordial
baryon-to-photon number
fluctuations are constrained by the observed structure of the
universe and by observationally inferred primordial light element
abundances. Fluctuations are also constrained by the high degree of
isotropy of the cosmic microwave background radiation (hereafter,
CMBR).
However,
we will assume here that an early reionization can erase any
preexisting anisotropies in the CMBR caused by the baryon-to-photon
number
fluctuations. It is not at all obvious whether a given PIB spectrum
of
fluctuations will result in early reionization and thus have a chance
of
escaping constraints from CMBR anisotropy limits.

The variance, or root-mean-square fluctuations in mass,
$(\delta M/M)_{\rm r.m.s.}$ , at the
present epoch has been determined by Davis \& Peebles (1983).
They find $(\delta M/M)_{\rm r.m.s.}\approx 1$ in a volume of size
$(8\, {\rm h^{-1} Mpc})^3$. This volume would correspond to a mass
scale of
$M_8\approx 6\times 10^{13}M_{\odot}(\Omega_bh^2/0.1)h^{-3}$.
It is well known that sub-horizon, super-Jeans-mass size baryon
number perturbations grow
proportionally  to the scale factor of the universe during a matter
dominated epoch. (cf. Kolb \&
Turner 1990).
This result generally obtains after recombination and when
fluctuations are in
the linear regime ($\delta M/M<<1$).
For example, such baryon number perturbations will have grown
by a factor of $(1+z_R)$ between the epoch of recombination and the
present epoch if a standard recombination scenario with recombination
redshift $z_R\approx 1100$ is assumed.

If the universe stays ionized at redshifts $z<z_R$, the growth of
perturbations will be inhibited by the coupling of photons to
baryons.
In the limit where the photon mean free path,
$l_{\gamma}$ , is larger than the fluctuation size, $l_f$ , growth of
baryon number perturbations is inhibited by Thomson drag
between photons and electrons. This drag force effectively
suppresses any perturbation growth for redshifts above
$z_{drag}\approx 200-300$ (cf. Peebles 1971). For redshifts below
$z_{drag}$,
Thomson drag rapidly becomes inefficient and so cannot hinder the
growth of
perturbations. In the limit when $l_{\gamma}<l_f$, sub-horizon
scale entropy
perturbations behave like oscillating sound waves.
Note that the photon mean free path has a comoving size of roughly
$\sim 8 {\rm h^{-1} Mpc}$ (corresponding to $M_8$) at a redshift of
$z\approx 200$.

In order that cosmic structure in
primordial isocurvature baryon number fluctuation models
not be \lq\lq overproduced\rq\rq\ , it is necessary that the
primordial root-mean-square mass fluctuations on the mass scale
$M_8$ be smaller than $(1+z)^{-1}$,
$$\sigma_{\eta}(M_8)\equiv\biggl({\delta M\over M}\biggr)_{M_8}
{}^<_{\sim}
{1\over 1+z}\ ,\eqno(16)$$
where $1100 {}^<_{\sim} z {}^<_{\sim} 200$.
Here the range in redshifts results from two extreme
scenarios: standard recombination; and ionization fully maintained
down to low
redshifts.

Note that equality in equation (16) pertains to the case
when the primordial isocurvature baryon number fluctuations
are the seeds for the presently observed cosmic structure.
It is, however, also conceivable that
the formation of structure on large scales is mainly due to, for
example, adiabatic fluctuations. In this case perhaps only the
small-mass scale fluctuations
are dominated by the primordial isocurvature baryon number
fluctuation component.
The inequality in equation (16) would apply in this latter example.

By using equations (15) and (16) we can derive a limit on the mass
cutoff,
$$M_c\, {}^<_{\sim}\, 6\times 10^{13}M_{\odot} \biggl({\Omega_bh^2
\over 0.1}\biggr)h^{-3}\Bigl\{(1+z_R)\sigma_{\eta}(M_c)
\Bigr\}^{-{6\over 3+n}}.\eqno(17)$$
Our nucleosynthesis results show that this mass cutoff must not fall
below $M_J^b$. If it did, we would produce unacceptable light element
abundances. Furthermore, unless $\sigma_{\eta}(M_c)\, ^>_{\sim}\, 1$
there
would be essentially no interesting effects on nucleosynthesis
(e.g., no early collapse of overdense regions with significant mass
fractions).
For a given spectral index it is not always clear that these two
requirements
are not mutually exclusive.

As an example of a scenario that {\it does} meet both criteria
consider
$M_c\approx 3\times 10^9M_{\odot}$, $\sigma_{\eta}(M_c)\approx 1$,
$n=0$,
and $z_r \leq 250$. By contrast, an example of a scenario which will
not
work has $M_c\approx 1\times 10^3M_{\odot}$,
$\sigma_{\eta}(M_c)\approx 10$,
$n=-1$, and $z_r\approx 1100$.

\vskip 0.15in
\centerline{\bf 2.4. Convergence of numerical results}
\vskip 0.1in

We have performed detailed primordial nucleosynthesis calculations
with
the numerical techniques outlined above. In Figures 4a and 4b we
present several
measures of convergence accuracy as functions of numbers of Fourier
modes
employed in our numerical study.

For the results shown in Figure 4a we employ $<\eta >=6\times
10^{-10}$,
$\Delta_{cr}=1.5$, $n=0$, and the transformation in equation (8a).
In Figure (4a) the upper panel shows the ratios of $^2$H/H (solid
line),
$Y_p$ (dotted line), and $^7$Li/H (short-dashed line) to their
converged values
as functions of the number of zones employed. Also shown in this
figure is the
ratio of $\Omega_b/\Omega_{diff}$ to its converged value. Here
$\Omega_{diff}$
is the fractional contribution of baryons in surviving, uncollapsed,
regions to
the closure density. It is clear that good convergence is obtained
for cases
with more than a few thousand Fourier modes.

The lower panel of Figure 4a shows $\sigma_{\eta}(M_c)$ (solid line)
and
$20\times \sigma_{\eta}(1000\times M_c)$ (dotted line) as functions
of the
number of Fourier modes. Convergence of $\sigma_{\eta}$ is good
whenever more
than a few thousand Fourier modes are employed.

For the computations presented in Figure 4b we employ $<\eta
>=1.2\times
10^{-9}$, $\Delta_{cr}=2.0$, $n=0$, and the transformation in
equation (8b).
The quantities plotted in these figures are the same as in Figure 4a,
and
the notation is the same. Again, we note the rapid convergence of
light element
abundance yields and $\Omega_b/\Omega_{diff}$ with increasing number
of
Fourier modes (upper panel). In the lower panel, however, we must
conclude
that the calculation for $\sigma_{\eta}$ remains unconverged, even
for
$10^5$ Fourier modes.

The reason for the lack of convergence in $\sigma_{\eta}$ for the
case in Figure
4b stems from the fact that the calculations of $\sigma_{\eta}$ are
dominated
by (rare) high-$\sigma$ events in the gaussian random variable. Note
that this
is not the case for the calculations of the light element abundances.
Therefore, we can predict the light-element abundances in this case
with
confidence, though we are unable to predict accurately the effective
spectral
index, $n$ , which correspond to our distribution for $\eta$ in the
cases where
the parameters of Figure 4b are adopted. Similar conclusions would
obtain
if we had employed the transformation law in equation (8c) instead of
that
in equation (8b).

\vskip 0.15in
\centerline{\bf 3. Results}
\vskip 0.1in

In this section we describe the results of our numerical study of
primordial
nucleosynthesis in the presence of large mass scale, PIB-like,
entropy
fluctuations. Here we will discuss not only the light element
($^2$H, $^3$He, $^4$He, $^7$Li) nucleosynthesis yields from various
models
for the distribution of $\eta$, but also the fraction of baryons in
these models
which survive collapse and comprise the \lq\lq diffuse\rq\rq\
primordial gas.
We will characterize the surviving diffuse component of baryons by
its
fractional contribution to closure, $\Omega_{diff}h^2$. In what
follows
$\Omega_bh^2$ refers to the baryonic fraction of the closure density
{\it prior} to freeze-out from nuclear statistical equilibrium and,
thus,
prior to any significant amount of collapse.

In Figure 5a and Figure 5b we present nucleosynthesis results for a
model of
the distribution of $\eta$ which is characterized by the
transformation
in equation (8a), spectral index $n=0$ (employed in equation 7),
$M_c>M_J^b$ , and various values of $\Omega_bh^2$ and $\Delta_{cr}$.
In Figure 5a the panel at upper left gives the $^4$He mass fraction,
$Y_p$,
as a function of $\Omega_bh^2$. The panel at upper right in this
figure
gives the ratio of produced $^2$H to hydrogen, $^2$H/H, as a function
of
$\Omega_bh^2$. Similarly, $^7$Li/H and $^3$He/H versus $\Omega_bh^2$
are
shown in the panels at lower left and right, respectively. In Figure
5b
we plot $\Omega_{diff}h^2$ versus $\Omega_bh^2$ for various models.
In Figure 5a the solid lines give the results for standard HBBN at
the indicated
values of $\Omega_bh^2$.

The other lines in Figures 5a and 5b are as follows: The dotted line
is for
$\Delta_{cr}=1.25$; the short-dashed line is for $\Delta_{cr}=1.5$;
the
long-dashed line is for $\Delta_{cr}=1.75$; while the dashed-dotted
line is
for $\Delta_{cr}=2.$

Note that several general trends are evident in Figures 5a and 5b.
First we note
that for all values of $\Delta_{cr}$ and $\Omega_bh^2$ considered
here $Y_p$
is lower than the HBBN yield, while $^2$H/H and $^3$He/H are higher
than the
HBBN yield. Our models clearly retain the well known \lq\lq $^7$Li
dip\rq\rq\ ,
but in general $^7$Li/H can be slightly smaller, comparable to, or
larger
by a factor up to 5, than the yield from HBBN at a given
$\Omega_bh^2$

These results are easily understood by comparison of the panels in
Figure 5a
to the graph in Figure 5b. Clearly $^2$H/H is high in inhomogeneous
models
compared to HBBN at the same $\Omega_bh^2$ because the surviving
diffuse baronic
component is characterized by $\Omega_{diff}h^2<\Omega_bh^2$. In
fact, since in
HBBN $^2$H/H yields rise very steeply with {\it decreasing} $\eta$ ,
we can identify two competing effects in $^2$H production. Regions
with smaller
$\eta$ produce relatively more $^2$H, but they make a relatively
smaller
contribution to the total surviving mass of baryons. We find that the
regions
which are most effective in producing $^2$H have $\eta\approx 3\times
10^{-11}$.
Since in primordial nucleosynthesis $^3$He is produced by
$^2$H(p,$\gamma$)$^3$He,
it is not surprising that $^3$He, like $^2$H, is always high compared
to
HBBN at the same $\Omega_bh^2$.

The situation for $^4$He is straightforward. Since $Y_p$ is a rising
function of $\eta$  in HBBN, it is obvious that when
$\Omega_{diff}h^2<\Omega_bh^2$
inhomogeneous models will produce low $Y_p$ relative to HBBN at the
same
$\Omega_bh^2$.

By contrast, the behavior of the $^7$Li/H yield in our inhomogeneous
models
is more complicated. In part, this is due to the two principle
production
channels for $^7$Li. These are $^3$H($\alpha$,$\gamma$)$^7$Li, which
is dominant
in low density regions where $^2$H (thus also $^3$H) is high, and
$^3$He($\alpha$,$\gamma$)$^7$Be($e^{-}$,$\nu_e$)$^7$Li, which is
dominant in
higher density regions. Since, to some extent, our models average
over regions
which have baryon-to-photon ratios on opposite sides of the dip, our
$^7$Li/H yields are high compared to HBBN for a fair range of
$\Omega_bh^2$.
Note, however, two interesting features of our $^7$Li/H results: (1)
the dip
is offset to higher $\Omega_bh^2$ than the HBBN dip; and (2) $^7$Li/H
at
higher $\Omega_bh^2$ is {\it lower} than the yield in HBBN.

A surprising feature of Figure 5a is that the light element abundance
yields
in our inhomogeneous model are actually fairly insensitive to the
values
of the parameter $\Delta_{cr}$ which is employed. We conclude that,
even in
cases where both the criterion in $\Delta_{cr}$ required for collapse
and the
collapse efficiencies are not accurately determined, we can predict
the
nucleosynthesis for a given PIB-like fluctuation spectrum with fair
confidence. This result is quite important since the collapse
efficiencies
may well depend not only on the fluctuation's overdensity but also on
internal
geometry, intrinsic angular momentum, and the local environment
(e.g., neighboring
fluctuations). We therefore expect collapse efficiencies to not
change
discontinously from zero to one at some value $\Delta_{cr}$, but
rather
approach unity in a continous fashion over some interval of
overdensities.

However, these conclusions are dependent on the
assumption that collapsing regions do not explode and produce large
amounts
of explosive nucleosynthesis ({\it cf.}, Fuller, Woosley, \& Weaver
1986).
Furthermore, the results in Figure 5 depend
critically on having a significant fraction
of the baryons collapse.
In the case where only a small fraction of the baryons collapse (i.e.
$\Delta_{cr}$ very large) we would find $Y_p$ and $(^7$Li/H) to be
increased
considerably over the results displayed in Figure 5a.

At this point we wish to summarize the observational constraints on
the
light-element abundances. The situation for the $^4$He mass fraction,
$Y_p$ , is as follows. The best determination of $Y_p$ is obtained
from
observations of $^4$He-recombination lines in metal-poor,
extragalactic
H II regions. The existing data has been analyzed by a number of
authors
(Walker {\it et al.} 1991; Fuller, Boyd, \& Kalen 1991; Pagel {\it et al.}
1992; Mathews, Boyd, \& Fuller 1993, Olive \& Steigman 1994).
It is generally believed that the upper
limit on $Y_p$ should be
somewhere in the range $Y_p {\
\lower-1.2pt\vbox{\hbox{\rlap{$<$}\lower5pt\vbox{\hbox{$\sim$}}}}\ }
0.24-0.245$, with the
\lq\lq favored\rq\rq\ value for $Y_p$ around $Y_p\approx 0.23$.

However, more recently it has been pointed out that the
observational determination
of $Y_p$ is subject to several systematic uncertainties which have
previously
not been well appreciated (Skillman \& Kennicut 1993; Skillman, Terlevich, \&
Garnett 1994; Sasselov \& Goldwirth 1994).
When all the systematic uncertainties are
taken
into account, such as uncertainties associated with the determination
of
emissivities and corrections due to the possible existence of some
neutral
helium in H II regions, a firm upper limit on $Y_p$ may be as large
as
$Y_p  {\
\lower-1.2pt\vbox{\hbox{\rlap{$<$}\lower5pt\vbox{\hbox{$\sim$}}}}\ }
0.252$. A lower limit on $Y_p$, which is of less
constraining power for PIB-like models than the upper limit, should
be
somewhere around $Y_p  {\
\lower-1.2pt\vbox{\hbox{\rlap{$>$}\lower5pt\vbox{\hbox{$\sim$}}}}\ }
0.21-0.22$.
It is clear that an accurate determination of the primordial
$^4$He-mass
fraction from the existing data is actually not that straightforward.

The situation is not very different for the primordial abundances of
$^2$H, and $^3$He. It is common practice to give the upper limit for
the
sum of $^2$H and $^3$He (Walker {\it et al.} 1991). This is done
since it is
not known to what extent $^2$H, the most fragile of the light
elements,
has been destroyed in stars prior to the formation of the solar
system.
Before the very recent claims of an extragalactic observation of
$^2$H by
Songaila {\it et al.} (1994), the deuterium-to-hydrogen number ratio
had
been estimated only for the solar system. By considering the sum of
$^2$H and
$^3$He some of the uncertainties in the determination of the
individual
abundances are evaded, since the destruction of $^2$H is expected to
lead
to the production of $^3$He via $^2$H(p,$\gamma$)$^3$He.
In any case, it is commonly believed that the upper limit on the sum
of these
light elements is ($^2$H+$^3$He)/H $^<_{\sim} 10^{-4}$
(Smith {\it et al.} 1993). The lower
limit for
deuterium is usually given at ($^2$H/H) $\simge 1.8\times 10^{-5}$.

Recently, Songaila {\it et al.} (1994) observed an isotope-shifted
Lyman-$\alpha$ absorption line in a Lyman-$\alpha$ cloud system which
lies
along the line of sight to  a quasar. The existence of an absorption
line is
well explained if the Lyman-$\alpha$ cloud has a
deuterium-to-hydrogen
number fraction as large as $(^2{\rm H/H})\approx 1.9\times
10^{-4}-2.5\times
10^{-4}$. This observation may cast doubt on the previously believed
upper
limit on primordial ($^2$H+$^3$He)/H. It may, however, also be that
the
detected absorption line comes from hydrogen as opposed to deuterium.
This could occur if, by coincidence, a small component of the
Lyman-$\alpha$
cloud system has a
small collective-velocity relative to the main cloud. In such
a case,  the existence of an
isotope-shifted Lyman-$\alpha$ absorption line might be mimicked.
Further
observations of Lyman-$\alpha$ cloud systems will hopefully provide
us
with a reliable primordial deuterium abundance.

The situation for the primordial $^7$Li abundance is controversial as
well.
Spite \& Spite (1982) detected a $^7$Li abundance plateau for hot,
low-metallicity Population II halo stars over some temperature and
metallicity range. A commonly held view
is that the $^7$Li-abundance
of $(^7{\rm Li/H})\approx 1.4\times 10^{-10}$ observed in the
\lq\lq plateau\rq\rq -stars is the primordial one
(Spite \& Spite 1982; Thorburn 1994). When uncertainties in the
employed
model atmospheres for the plateau-stars and \lq\lq small\rq\rq\
amounts of
diffusion-induced $^7$Li depletion are taken into account, the upper
limit
on the primordial value of  $^7$Li may be larger by a factor which is
roughly between
one and two
(Deliyannis, Demarque, \& Kawaler 1990).

The above interpretation of the data precludes the possibility
of significant $^7$Li depletion in plateau-stars. However, it has
been shown
by Pinsonneault, Deliyannis, \& Demarque (1992) and Chaboyer \& Demarque (1994)
that $^7$Li
in Population II stars is depleted by up to an order of magnitude, or
more,
over the entire surface temperature range (including the
Spite-plateau) when
microscopic diffusion and rotation of stars are included in the
stellar models.
Furthermore, only combined models which include rotation and
diffusion
can simultaneously explain the very different $^7$Li depletion
patterns
in old Population II stars and young Population I stars
(Pinsonneault {\it et al.} 1992; Chaboyer, Demarque, \& Pinsonneault 1994ab).
This may be a powerful argument for the validity of the combined
models.
Combined models predict a $^7$Li-abundance of $(^7 {\rm Li/H})\sim
1\times
10^{-9}$, much larger than the value of the Spite-plateau.

It has been pointed out that a possible detection of $^6$Li in two
population II stars (Smith, Lambert, \& Nissen 1982; Hobbs \&
Thorburn 1994)
may provide an argument against significant depletion. This is
because
$^6$Li, presumbly produced by cosmic ray spallation of heavier
elements
over the entire history of the universe, would have been depleted
below
detection level along with $^7$Li. On the other hand, the
observations of
three highly $^7$Li-depleted plateau stars (Thorburn 1994) indicates
that significant $^7$Li-depletion in plateau stars may occur, at
least for some stars.

The situation remains controversial. It should be noted that the
primordial
$^7$Li abundance, when determined with confidence, could be a crucial
argument against, or indication for, the existence of small-scale
and/or large-scale
baryon-to-photon inhomogeneity during the nucleosynthesis epoch.

The implications of the observatially inferred primordial abundance
constraints for the model results displayed in Figures 5a and 5b are
as
follows. There seem to be two possible ranges in $\Omega_bh^2$ which
could
yield agreement with abundance constraints within the above-mentioned
observationally inferred primordial abundance uncertainties.
For $\oh\approx 0.015$ we find from Figure 5a that $\h2\approx
2\times 10^{-4}-4\times 10^{-4}$; $Y_p\approx 0.225-0.235$;
and $\li7\approx 3.5\times 10^{-10}-4.5\times 10^{-10}$ depending on
the
employed $\Delta_{cr}$. To be correct, this scenario would have to
assume that there was a modest amount of
$^7$Li depletion in halo stars and a high primordial deuterium
abundance.

For the range $\oh\approx 0.03-0.05$,
 we obtain $\h2\approx 4\times 10^{-5} - 1.5\times
10^{-4}$; $Y_p\approx 0.24-0.25$; and $\li7\approx 5\times 10^{-10} -
1.5\times 10^{-9}$. This model would then require a significant
$^7$Li depletion,
a relatively low deuterium abundance and high $^4$He abundance.
For comparison, the preferred range for $\oh$ in homogeneous big bang
nucleosynthesis is somewhere in the range $\oh\approx 0.005-0.015$,
mostly
depending on what primordial deuterium abundance is adopted.

Another generic feature of our models with large-scale inhomogeneity
in
$\eta$ is evident from Figure 5b. Even though the abundance
yields in these models tend to agree with observationally inferred
abundance
limits for higher $\oh$ than the allowed range of this quantity
in standard homogeneous big bang
models, the fractional contribution of the diffuse, \lq\lq
surviving\rq\rq\
baryons to the closure density, $\Omega_{diff}h^2$, tends to be {\it
lower} than the
allowed range of $\Omega_bh^2$ in
homogeneous models. For a good agreement between
abundance yields and observationally inferred abundance limits in
PIB-like
models, the low-density regions must
have approximately the preferred average
standard homogeneous baryon-to-photon ratio (i.e. $\oh\approx 0.0125$
for $\h2\simle 10^{-4}$). Therefore, $\Omega_{diff}h^2$ is lower than
the $\oh$
from a homogeneous big bang, since only the low-density regions in
the
universe contribute to the diffuse baryons. The high-density regions
do not
contribute since these presumbly collapse and form condensed, dark
objects.

This way of lowering $\Omega_{diff}h^2$ compared to $\oh$ inferred
from
homogeneous big bang nucleosynthesis is quite analogous to the
scenario
proposed by Jedamzik, Mathews, \& Fuller (1994). In that paper it was
shown
that there is no lower limit on $\oh$ in inhomogenous primordial
nucleosynthesis scenarios, since it may be that only a certain
fraction of space
has baryons with baryon-to-photon ratio $\eta\approx 3\times
10^{-10}$,
with the remaining fraction of space depleted in baryons.
Such a scenario could provide an explanation for the small amount of
baryons
observed in luminous form (i.e., galaxies and diffuse intergalactic
gas).
The fractional contribution of luminous baryons to the closure
density is
$\Omega_b^{lum}\approx 0.003-0.007$. This value can be much smaller
than
the $\Omega_b$ inferred from homogeneous big bang nucleosynthesis. It
is
interesting to note that inhomogeneous PIB-like models conceivably
could
provide a natural explanation for the small value of
$\Omega_b^{lum}$.

In Figure 6a and 6b we present nucleosynthesis results for a model of
the
distribution of $\eta$ which is characterized by the transformation
in equation (8b), spectral index $n=0$, $M_c>M_J^b$, and various
values of
$\oh$ and $\Delta_{cr}$. The notation is as in Figure 5a and 5b. The
assumed values for $\Delta_{cr}$ in these calculations are as
follows: the
dotted line is for $\dc =1.5$; the short-dashed line is for $\dc =2$;
the long-dashed line is for $\dc= 2.5$; and the dashed-dotted line is
for
$\dc =3$.

It is seen from Figure 6a and 6b that most of the general trends
observed
in the PIB-like model of Figure 5a and 5b are retained. It is evident
that
there is a range in $\oh$ ($0.04 \simle \oh \simle 0.07$) for which
all the
abundance constraints may be satisfied. In this range we find
$\h2\approx 9\times 10^{-5}-3\times 10^{-4}$; $Y_p\approx
0.235-0.248$;
and $\li7\approx 7\times 10^{-10}-2\times 10^{-9}$, depending on the
value of
 $\dc$ which is employed. Clearly, such models would be ruled out if
the
$^7$Li abundance
of the Spite-plateau in halo stars represents the actual primordial
abundance. Low values for $\oh\simle 0.02$ are ruled out by deuterium
overproduction. While the models of Figure 5 are characterized by
collapse
ratios $\Omega_b/\Omega_{diff}\approx 2.5-10$, a much larger fraction
of the
baryons would collapse in the model of Figure 6,
$\Omega_b/\Omega_{diff}\approx
25-75$.

In Figure 7a and 7b we present the results for a distribution in
$\eta$
caracterized by the transformation in equation (8c), spectral index
$n=0$
and $M_c>M_J^b$. As in Figure 5 and 6 we give results for various
$\oh$
and $\dc$. The lines are for $\dc =1.5$ (dotted line), $\dc =2$
(short-dashed line), $\dc =2.5$ (long-dashed line), and
$\dc =3$ (dashed-dotted line).

In this model, the low $^4$He-mass fraction, and high $\h2$-number
ratio
compared to a HBBN model at the same $\oh$ is most pronounced. The
collapse
ratios, $\Omega_b/\Omega_{diff}$, are very similiar to the collapse
ratios in
the model of Figure 6, in particular $\Omega_b/\Omega_{diff}\approx
25-75$.
In the range $\oh\approx 0.06-0.2$ we obtain abundance yields of
$\h2\approx 4\times 10^{-5}-4\times 10^{-4}$, $Y_p\approx
0.225-0.25$,
and $\li7\approx 9\times 10^{-10}-3\times 10^{-9}$ depending on the
value
for $\dc$. These abundance yields may agree with observationally
inferred
abundance limits when significant $^7$Li depletion in population II
halo
stars occurs. It is remarkable that the allowed (albeit with high
$^7$Li)
range of $\oh$ in these models is
between a factor of ten and fourty larger than the allowed range of
$\oh$ in HBBN
models. The upper end of this range in $\oh$ may even allow for
baryons
to provide closure density when the Hubble parameter is smaller than
$h\simle 0.45$.

We have also investigated the nucleosynthesis results in PIB-like
models as
a function of the collapse efficiency parameter $f$ as defined in
Section 2.
In Figures 8a and 8b we present the nucleosynthesis yields and ratios
$\ooo$
for a distribution in $\eta$ characterized by the transformation in
equation
(8a), spectral index $n=0$, $M_c>M_J^b$, and $\dc=1.5$. In this
figure
we vary $\oh$ and the collapse efficiency parameter. The notation in
these
figures is similiar to that in Figures 5,6 and 7.

The different lines correspond to the following collapse
efficiencies:
$f=100\%$ (lower dotted lines), $f=97.5\%$ (short-dashed lines),
$f=95\%$
(long dashed lines), $f=90\%$ (dashed-dotted lines), and $f=80\%$
(upper dotted lines). Note that the lower dotted lines in Figures 8a
and 8b
represent the results for the same model as the short-dashed lines
in Figures 5a and 5b. These lines
are shown for comparison. Note
that the lower dotted lines in the $Y_p$ and $\li7$ panels of Figures
8a
and 8b are to be associated with the upper dotted lines
 in the $\h2$ and $\he3$ panels of Figure 8a.

It is not surprising to find that the $^4$He and $^7$Li abundances
increase
and the $^2$H and $^3$He abundances decrease when the collapse
efficiency
parameter decreases. This is because with lower collapse efficiency
more
high-density regions contribute their nucleosynthesis yields to the
diffuse
baryons. The high-density regions produce relatively larger amounts
of $^4$He
and $^7$Li and smaller amounts of $^2$H and $^3$He than the
low-density
regions. In this model, however, it is evident that collapse
efficiencies do
not have to be extremely close to 100\% in order
to avoid gross overproduction of
$^4$He and $^7$Li.

In Figures 9a and 9b we present the nucleosynthesis yields and
survival
ratios $\ooo$
for a distribution in $\eta$ characterized by the transformation
equation
(8b), spectral index $n=0$, $M_c>M_J^b$, and $\dc =2$. We show models
with
collapse efficiencies $f=100\%$ (dotted lines), $f=99\%$
(short-dashed
lines), and $f=97.5\%$ (long-dashed lines). Note that the dotted
lines in
Figures 9a and 9b represent the results for the same model as the
short-dashed lines in Figure 6a and 6b.

{}From Figures 9a and 9b it is evident that the
collapse efficiencies for these models
would have to be very close to $100\%$
in order to avoid
overproduction of
 $^4$He and, especially, $^7$Li.
This conclusion stands in contrast to that derived
from the models of Figures
8a and 8b, and is easily understood by
an examination of the results
presented in
Figure 1b. This figure shows a small part of the distribution in
$\eta$ from which the model results of Figures 9a and 9b have been
computed.
The distribution is characterized by very overdense peaks. If only a
very
few of these peaks did not collapse, their contributions to the
$^7$Li component in the surviving
diffuse baryons from these high-density regions
would be very significant, perhaps even dominant.

Our computation of the abundance yields from
a particular model as a function of the
collapse efficiency parameter $f$ proceeds in the following
manner. First, the algorithm finds all the regions which are
overdense by the
critical amount $\dc$. Then, the algorithm randomly chooses a
fraction
$(1-f)$ of these overdense regions to not collapse and thus to
contribute to the abundance yields of the surviving diffuse baryons.

This procedure may actually not lead to a very accurate assessment of
the
variations in abundance yields associated with the uncertainties of
collapse
of particular regions. Realistically, an overdense region may not
completely
collapse because of its pecuilar angular momentum and local
environment.
However,
we expect the very overdense peaks within an overdense region to
collapse
with high efficiency. A better assessment of the variations in
abundance
yields due to collapse efficiency uncertainties may be obtained by
averaging
over the abundance yields for a range of $\dc$. Figures 5-7, which
show the
abundance yields of different models for different $\dc$, may
therefore
provide a more realistic estimate for the anticipated magnitude of
variations in abundance yields due to collapse efficiency
uncertainties. In
any case, Figure 9a illustrates that a very large fraction ($\simge
99\%$)
of the overdense peaks of a distribution characterized by very
overdense peaks
must collapse in order that $^4$He and $^7$Li not be overproduced.

In Figure 10a and 10b we show the nucleosynthesis results and ratios
$\ooo$
for a distribution in $\eta$ characterized by transformation equation
(8c),
spectral index $n=0$, $M_c>M_J^b$, $\dc =2$, and varying $\oh$ and
collapse
efficiencies $f$. The lines represent $f=100\%$ (dotted lines),
$f=99\%$
(short-dashed lines), and $f=97.5\%$ (long-dashed lines). The
conclusions
drawn from these figures are quite similiar to the conclusions for
the model
investigated in Figures 9a and 9b. A large fraction ($f \simge 99\%$)
of
very overdense peaks has to form dark remnants in order that $^7$Li
and $^4$He
not be overproduced.

In our study we have so far assumed that the fluctuation cutoff mass
scale,
$M_c$, is larger than the mass scale $M_J^b$. The local baryon
Jeans mass, $M_J^b$, divides
fluctuation evolution into two regimes: overdense fluctuations on
mass scales
$M>M_J^b$ ultimately are expected to collapse; whereas, overdense
fluctuations
on mass scales $M<M_J^b$ will expand, and their nucleosynthesis
yields will
mix with the diffuse baryons. We have examined the nucleosynthesis
yields of
PIB-like models when fluctuations exist on mass scales below $M_J^b$.

In Figures 11a and 11b we show the nucleosynthesis yields and
fraction $\ooo$
for a PIB-like model with a distribution in $\eta$ characterized by
the
transformation equation (8a), spectral index $n=0$, $\dc =1.5$, and
collapse
efficiency $f=100\%$. In this figure we have varied the cutoff mass
scale
$M_c$. The dotted line is for $M_c>M_J^b$, the short-dashed line is
for
$M_c=M_J^b/3$, the long-dashed line is for $M_c=M_J^b/6$, and the
dashed-dotted line is for $M_c=M_J^b/10$. Note that the dotted lines
represent the results of our \lq\lq reference\rq\rq\ model, which
have already
been shown by the short-dashed lines in Figure 5a and 5b and the
dotted lines
in Figure 8a and 8b.

Compared to the reference model, the $^4$He and $^7$Li yields
increase and
the $^2$H
and $^3$He yields decrease
whenever $M_c$ falls below $M_J^b$. These results are
as expected, since sub-Jeans mass size overdense fluctuations which
are not
included in super-Jeans mass size overdense regions will expand and
contribute large amounts of $^4$He and $^7$Li to the diffuse baryons.
This
material from sub-Jeans mass size overdense fluctuations will also
dilute the
$\h2$ and $\he3$ ratios relative to the reference model. Note that
$^4$He and
$^7$Li (for most of the \lq\lq interesting\rq\rq\ range in $\oh$)
will even
increase over the HBBN yields as the fluctuation cutoff mass scale is
decreased.

It is evident from Figure 11b that
the fraction of material included in regions which meet
the requirement to have overdensity $\dc$
on a mass scale larger than $M_J^b$ decreases as
the cutoff mass scale $M_c$ decreases. Equivalently,
the ratio $\ooo$ decreases with decreasing
$M_c$ as well. In the limit where $\ooo$
approaches unity, there is no collapse and the abundance yields would
be given
by a weighted average over the HBBN yields of all the regions of the
entire
distribution. The model results displayed by the dashed-dotted lines
are
actually not far removed from that limit ($\ooo \simle 2$).
For comparison, the same model but {\it without} any collapse ($\ooo
=1$)
would yield $\h2 =1.5\times 10^{-4}$, $Y_p=0.242$, and $\li7
=8.4\times
10^{-10}$ for $\oh =0.01$. These abundance yields differ only
slightly
from the yields for $\oh =0.01$ (dashed-dotted line).

It is quite surprising that even for a distribution of $\eta$ as
inhomogeneous as that in Figure 1a, and without any collapse, there
exists
a narrow interval in $\oh$ where all the observational abundance
constraints
may be satisfied. In particular, for $0.006 \simle \oh \simle 0.0012$
agreement between the computed abundance yields and the
observationally
inferred abundance constraints is possible within the observational
uncertainties. Note that this range for $\oh$ is close to the range
for $\oh$
inferred from HBBN. Agreement between abundance yields and
observationally
inferred abundance limits would require significant $^7$Li depletion
and a
high deuterium abundance $\h2 \simge 1\times 10^{-4}$.

We have implicitly assumed so far that there is a cutoff mass scale
which is
larger than $M_c\simge 10^{-11}M_{\odot}$ (approximately sixteen
orders of
magnitude smaller than $M_J^b$!). If there were
fluctuations on mass scales below $M\simle 10^{-11}M_{\odot}$,
diffusive and
hydrodynamic processes during the epoch of primordial nucleosynthesis
would
alter the abundance yields from what we have calculated.

It is not straightforward to estimate accurately the effects of
diffusive
processes during the nucleosynthesis era on the abundance yields of
{\it
stochastic} baryon number distributions such as those shown in
Figures 1abc.
The nucleosynthesis yields of a regular lattice of fluctuation sites
in the
mass range $10^{-21}M_{\odot}\simle M \simle 10^{-11}M_{\odot}$ have
been
investigated in detail (Kurki-Suonio {\it et al.} 1988; 1990; Mathews {\it
et al.} 1990; Jedamzik {\it et al.} 1994).
Depending on the fluctuation characteristics,
especially the separation of adjacent fluctuations in the regular
lattice,
$Y_p$ is
usually larger (but can be smaller) than the $Y_p$ of a homogeneous
model at
the same average $\oh$. The number ratio of $\li7$ tends to be much
larger in
such inhomogenous models than its HBBN value. On the other hand,
 most inhomogeneous scenarios {\it
with} diffusion tend to yield slightly lower values for $Y_p$ and
$\li7$
than do
inhomogeneous models which neglect diffusive processes.

The abundance yields for a truly stochastic distribution may be
roughly
approximated by averaging over the abundance yields from different
regular
lattices of fluctuation sites with varying fluctuation separation
lengths.
Such an average should always increase $Y_p$ and $\li7$ relative to
their
respective HBBN yields (Meyer {\it et al.} 1991).
Therefore, we expect models which have small-scale
fluctuations and diffusive processes during the nucleosynthesis era
to produce
$^4$He and $^7$Li yields which are
far above the dotted lines shown in Figure 11a.
We also expect such models to yield $^7$Li and $^4$He slightly
below the dashed-dotted lines in Figure 11a. The situation for the
other
light elements, $^2$H and $^3$He, is more complicated.
These issues are investigated in more detail by Kurki-Suonio,
Jedamzik, \&
Mathews (1994). These authors will treat diffusive processes during
the
nucleosynthesis epoch explicitly and combine their results with the
results
of the present study.

Note that our arguments implicitly assume that there is some turnover
in
effective spectral index of the baryon number distribution, from
$n_{eff}=0$
on large mass scales, to an $n_{eff}=-3$ Harrison-Zeldovich
character on small mass scales.
If there were not such a turnover in effective spectral index, there
would either be no significant fraction of baryons collapsing (since
fluctuation amplitudes on $M_J^b$-size mass scales are very small),
or
fluctuation amplitudes on small mass scales $M\simle
10^{-11}M_{\odot}$ would
have to be extremely large ($\Delta \simge 10^8$).

In Figures 12a and 12b we show the nucleosynthesis yields and
fraction $\ooo$
for a distribution of $\eta$ characterized by the transformation
equation
(8b), spectral index $n=0$, $\dc =2$, and varying $\oh$ and $M_c$.
The lines
represent results for $M_c> M_J^b$ (dotted), $M_c=M_J^b/3$
(short-dashed),
$M_c=M_J^b/6$ (long-dashed), and $M_c=M_J^b/10$ (dashed-dotted). In
these
figures the dotted line represents the results of our reference model
which
have already been shown by the short-dashed lines in Figures 6a and
6b and
the dotted lines in Figures 9a and 9b.

When the cutoff mass scale is decreased, $Y_p$ and $\li7$ abundance
significantly
increase, and $\h2$ and $\he3$ decrease relative to the results of
the
reference model. Since this particular distribution in $\eta$
includes very
overdense peaks ($\Delta\sim 100$), it is not surprising that the
$^4$He and
$^7$Li yields {\it increase} over the yields of a HBBN scenario at
the same
$\oh$.

For the particular distribution investigated here, there seems to be
no
interval in $\oh$ for which all the abundance constraints may be met.
This
result implies that models with stochastic large- and small-scale,
large-amplitude
($\Delta\sim 100$) inhomogeneities in the baryon-to-photon ratio are
ruled
out by the observationally inferred primordial light element
abundances. We can only expect possible
agreement between the abundance yields
of such models and observationally inferred abundance limits when
there is a
fluctuation cutoff mass scale $M_c\simge M_J^b$. Furthermore, our
discussion
about the effects of diffusive processes during the nucleosynthesis
era
should indicate that the existence of small-scale fluctuations down
to the
mass range $M\sim 10^{-21}-10^{-11}M_{\odot}$ is not likely to change
these
conclusions.

In Figures 13a and 13b we present the nucleosynthesis yields from a
distribution in $\eta$ characterized by transformation equation (8c),
spectral index $n=0$, $\dc =2$, and for cutoff mass scales
$M_c>M_J^b$
(dotted), $M_c=M_J^b/3$ (short-dashed), $M_c=M_J^b/6$ (long-dashed),
and
$M_c=M_J^b/10$ (dashed-dotted). It is evident from these figures that
for
$M_c<M_J^b$ there seems to be no range in $\oh$ where the production
of all
the light elements is consistent with observationally inferred
abundance
constraints. Note that this model is characterized by large-amplitude
fluctuations, and so is similiar to the model considered in Figures
12a and 12b.

We have investigated the dependence of nucleosynthesis yields in
PIB-like
models on the effective spectral index of the distribution in $\eta$.
We have
employed spectral indices $n=0$ and $n=-2.4$ in equation (7). Note
that $n$
here denotes the three-dimensional spectral index. The
$\sigma_{\eta}(M)$
(or, equivalently, $\delta\rho /\rho$ as a function of mass scale)
for the
resultant distributions in $\eta$ has been presented in Figures 3abc.
These
figures illustrate that the effective spectral index of the
distribution in
$\eta$ is $n_{eff}\approx 0$ for $n=0$ and $n_{eff}\approx -1.5$ for
$n=-2.4$. We note, however, that there is some uncertainty associated
with
the exact value of $n_{eff}$.

In Figures 14a and 14b we show the nucleosynthesis yields and $\ooo$
for a
distribution in $\eta$ characterized by transformation equation (8a),
$\dc
=1.5$, $M_c>M_J^b$, and effective spectral indices $n_{eff}\approx 0$
(dotted
lines) and $n_{eff}\approx -1.5$ (dashed lines). It is surprising to
find
that there is so little dependence of the nucleosynthesis yields on
the
effective spectral index of the distribution in $\eta$. In general,
for
decreasing $n_{eff}$ the $^4$He mass fraction, $Y_p$, decreases
slightly,
whereas the number ratios $\h2$ and $\he3$ increase slightly. The
number
ratio $\li7$ decreases for large $\oh$, and increases for small $\oh$
when
$n_{eff}$ decreases.

These trends can be easily understood on inspection of Figure 14b.
For a
smaller $n_{eff}$, a larger fraction of the baryons collapse. In
turn, a larger
fraction of collapsing baryons implies that the average
$\Omega_{diff}h^2$ is smaller, which ultimately leads to smaller
$^4$He yields
and larger $^2$H and $^3$He yields. The $^7$Li yield increases or
decreases
for decreasing $\ooo$ depending on whether $\oh$ is smaller than,
or larger than, the
$\oh$ at the $^7$Li dip. Our results show that the nucleosynthesis
yields in
PIB-like models are not very dependent on the effective spectral
index of the
distribution of $\eta$ in the interval $-1.5\simle n_{eff}\simle 0$.
Note
that PIB-models for
large scale structure formation would prefer spectral indices
somewhere
between $n_{eff}=0$ and $n_{eff}=-1$.

These conclusions are confirmed by the results shown in
Figures 15a and 15b and Figures 16a and 16b. These figures show
nucleosynthesis
yields and $\ooo$ for models with $\dc =2$, $M_c>M_J^b$, and two
different
spectral indices: $n_{eff}\approx 0$ (dotted lines) and
$n_{eff}\approx -1.5$
(dashed lines). The calculations described by
Figures 15a and 15b employ transformation equation
(8b) for the generation of the distribution of $\eta$; whereas,
calculations described by
Figures 16a and 16b employ transformation equation (8c) for the
generation of the distribution of $\eta$.

There is a unique prediction of cosmological models which contain
non-linear, intermediate to large scale primordial isocurvature
baryon
number fluctuations. Since such models are characterized
by inhomogeneity in the baryon-to-photon ratio during the
nucleosynthesis
epoch, we expect the production of
{\it intrinsic} spatial variations in the primordial
light element abundances. In contrast, HBBN models predict a
universal set of
cosmic light element abundances.

To illustrate this point, we define the probability distribution
function
$P_{\lambda}(Y_p)$. This function gives the probability for finding a
region
of size $\lambda$ to have average $^4$He mass fraction $Y_p$. Note
that the
definition of $P_{\lambda}(Y_p)$ is analogous to the definition of
$P_{\lambda}(\eta )$ in Section 2. In a similar fashion we define
probability
functions $P_{\lambda}\h2$, $P_{\lambda}\he3$, and
$P_{\lambda}\li7$
to give the probability for finding a region of size $\lambda$ with
average
number ratios of $\h2$, $\he3$, and $\li7$, respectively.

In Figures 17a, 17b, and 17c we display these numerically determined
probability functions for various scales $\lambda$ and different
models of
the distribution of $\eta$. In these figures the panels in the upper
left-hand corners display $P_{\lambda}(Y_p)$, whereas the panels in
the lower
left-hand corners display $P_{\lambda}\li7$. The panels in the
upper and
lower right-hand corners display $P_{\lambda}\h2$ and
$P_{\lambda}\he3$,
respectively. In each panel we show four probability distribution
functions. These distribution functions are determined for the scales
$\lambda <\lambda_c$ (solid lines), $\lambda =100\lambda_c$ (dotted
lines),
$\lambda =1000\lambda_c$ (short-dashed lines), and $\lambda
=10000\lambda_c$
(long-dashed lines). Note that in our \lq\lq one-dimensional\rq\rq\
theory
the length scale $\lambda$ is proportional to the mass scale $M$.

In Figure 17a we show probability distribution functions for a model
spatial distribution
of $\eta$ which is characterized by the transformation equation (8a),
spectral index $n=0$, $\dc =1.5$, and $M_c>M_J^b$. Furthermore, we
assume a
cosmic average baryon-to-photon ratio of $<\eta >=6\times 10^{-10}$.
This
corresponds to a value of $\oh =0.0224$.

The panels in Figure 17a clearly illustrate that there is a finite
width
to the abundance probability distribution functions.
In other words, there is a finite probability to find
regions of mass scale $M$ to have abundances
which are smaller, or larger, than average cosmic
abundances. The widths of the distribution functions decrease as the
mass
scale (or, equivalently, $\lambda$) increases. On the smallest mass
scales
(solid lines), there is a fairly wide
range of about equally probable primordial
abundances. On the largest mass scales (long-dashed lines), the
intrinsic
widths of the probability distribution functions are approximately
$4\times 10^{-3}<Y_p >$ for the $^4$He mass fraction, $10^{-1}<\h2 >$
for the
deuterium-to-hydrogen number ratio, $5\times 10^{-2}<\he3 >$ for the
$^3$He-to-hydrogen number ratio, and $7\times 10^{-2}<\li7 >$ for the
$^7$Li-to-hydrogen number ratio. Here the brackets denote the cosmic
average
abundances for this particular model for the distribution of $\eta$.

In Figure 17b we show probability
distribution functions for a spatial distribution of
$\eta $ which is characterized by the transformation equation (8b),
spectral
index $n=0$, $\dc =2$, and $M_c>M_J^b$. In this model we assume a
cosmic
average baryon-to-photon ratio of $<\eta >=1.2\times 10^{-9}$. We
observe
the same trends and features in this figure as were observed in
Figure
17a. However, the intrinsic widths in the distribution functions for
the
largest scales ($\lambda =10^4\lambda_c$) are much larger than those
in
Figure 17a. This is because the distribution investigated in Figure
17b
includes large-amplitude fluctuations in $\eta$, whereas the
fluctuation
amplitudes of the distribution investigated in Figure 17a are
moderate. On
the largest scales we find approximate widths of the distribution
functions
to be $2\times 10^{-2}<Y_p >$ for the $^4$He mass fraction, $0.4<\h2
>$ for the
deuterium-to-hydrogen number ratio, $0.25<\he3 >$ for the
$^3$He-to-hydrogen
number ratio, and $0.3<\li7 >$ for the $^7$Li-to-hydrogen number
ratio.

In Figure 17c we display distribution functions for a spatial
distribution of
$\eta$ which is characterized by the transformation equation (8c),
spectral
index $n=0$, $\dc =2$, and $M_c>M_J^b$. For this calculation we have
assumed
a cosmic average baryon-to-photon ratio of $<\eta >=3\times 10^{-9}$.

For the smallest scales, the probability distribution functions
display a peak
which is not centered at the cosmic average abundances. This is
easily
understood by inspection of Figure 1c, where it is
obvious that in the present model
there exists a minimum baryon-to-photon ratio $\eta_{min}$. A large
fraction
of the universe in this model has $\eta_{min}$, so that the
probability to find a small
region with abundance yields pertaining to $\eta_{min}$ is large. The
intrinsic wdths of the distribution functions on the largest scales
are
somewhat smaller than those found in Figure 17b.

These results imply that PIB-like models which have an
intermediate-scale,
non-linear fluctuation component would lead to intrinsic primordial
abundance
variations on small, as well, as large mass scales. Note that
a mass scale of $M=10^4M_c$ assumed for
the probability distribution functions
shown in Figures 17abc can easily
correspond to mass scales as large as $M\sim
10^{11}-10^{12}M_{\odot}$,
depending on
the assumed value of the cutoff mass scale, $M_c$.
We do not expect small-scale primordial abundance variations to
survive to
the present epoch. This is because mixing mechanisms, such as shock
waves
induced by supernovae explosions, should be efficient enough to mix
the
primordial material on intermediate mass scales.

It is, however, questionable if mixing could erase preexisting
primordial
abundance variations on mass scales as large as $M\sim
10^9-10^{11}M_{\odot}$. In fact, it is well known that the $^4$He
mass
fraction observed in metal-poor, extragalctic H II regions exhibits a
significant variation between
objects of the same
metallicity. Typical $^4$He mass fractions are in the range
$0.22-0.26$. It
has been suggested that this spread is not due to observational
uncertainties,
but rather represents a real physical variations.
However, even if these suggestions
are correct, there are several chemical evolution effects which may
produce
such a spread (Campbell 1992).

In any case, the intrinsic spread in the $^4$He mass fractions
observed in
metal-poor H II regions may be used to put upper limits on the
magnitude of
preexisting primordial abundance variations. In order to derive
useful upper
limits on large-scale inhomogeneity, the details of the mixing of
primordial material would have to be considered. In this way,
cosmological
models which include primordial, non-linear, isocurvature baryon
number
fluctuations could be further constrained.

\vfill\eject
\centerline{\bf 4. Conclusions}
\vskip 0.1in
We have examined the primordial nucleosynthesis
process in the presence of large mass
scale, nonlinear entropy fluctuations.  A variety of fluctuation
spectra were considered, including
some which have the characteristics of the spectra
of PIB models extrapolated to
smaller mass scales. Our computations provide
for the collapse of overdense regions with masses above
the local baryon Jeans mass, $M_J^b$ (equation 4).
The baryons, and hence the nucleosynthesis products,
which avoid incorporation into condensed objects
were found to originate mostly from underdense regions.
A complicating feature of nucleosynthesis calculations
in a field of stochastically distributed fluctuations arises
from the fact that a particular underdense region may reside inside a
larger overdense region which, in turn,
might be destined for collapse. In our computations we
have included a detailed numerical treatment of
this \lq\lq cloud-in-cloud\rq\rq\ problem,
and we have found that such a treatment was
important for the accurate estimate of light element
($^2$H, $^3$He, $^4$He, $^7$Li) abundance yields.

In general, we have found that PIB-like spectra with
significant small-scale, large-amplitude structure ({\it i.e.},
non-linear, $\Delta >>1$, structure on
mass scales $M < M_J^b$) produced light element
abundance yields which were in conflict with
observationally-inferred limits for almost any
pre-nucleosynthesis ${\Omega}_b$. This may
represent an important nucleosynthesis-based
constraint on PIB models for large scale structure formation,
if the underlying microscopic mechanisms
which generated fluctuations in these
models somehow
{\it demanded} the presence of nonlinear small-scale structure.

However, if such small scale fluctuations were absent or
suppressed, then our computations have
shown that there exists a range of
fluctuation spectral characteristics which produce
light element abundance yields
in agreement with observationally-inferred limits.
Particular distributions of baryons were found to produce acceptable
nucleosynthesis yields even when the pre-nucleosynthesis
baryonic fraction of the closure density was as high as
${\Omega}_b \approx 0.2 h^{-2}$ ({\it i.e.}, roughly closure density
for
small Hubble parameter).
On the other hand, the fractional contribution of the diffuse, \lq\lq
uncollapsed\rq\rq\ baryons to the closure density, $\Omega_{diff}$, was found
to tend to be lower than the $\Omega_b^{\rm HBBN}$ inferred from standard
homogeneous big bang nucleosynthesis. Typical values for $\Omega_{diff}<0.01$
may be in better accord with the observed fractional contribution of the
luminous baryons to the closure density ($\Omega_{lum}\approx 0.003-0.007$)
than $\Omega_b^{HBBN}$.

In any case,
we have found that such a relaxation of the
homogeneous big bang limit on ${\Omega}_b$ would
usually require that the primordial abundance of
 $^7$Li be closer to the observed Population I
value than to the Spite \lq\lq plateau\rq\rq\ value. In
turn, this would demand that there had been a fair amount of
destruction of $^7$Li in the Spite plateau stars, a conceivable
though
controversial possibility. Future observations may resolve this
issue.

In our models which met abundance constraints we found that
$^2$H/H was high and the $^4$He mass fraction low relative to
a homogeneous big bang at a given value of ${\Omega}_b h^{2}$.
Obviously, this would have to be the case if these models were
to meet abundance limits for values of ${\Omega}_b$ that were
larger than the limit on this quantity from the homogeneous big bang.
The important point is that PIB-like models possessing small scale
fluctuation cutoffs
could alter the relative abundances
of $^2$H, $^3$He, $^4$He, $^7$Li over
those from a homogeneous big bang,
while still meeting abundance constraints within observational
uncertainties. More accurate observational
determinations of any two of these
light element abundances might provide a signature for or,
more likely, a constraint on such PIB-like models.

We have pointed out that a potential signature
of a PIB-like distribution of entropy at the nucleosynthesis
epoch would be the observation of a significant and {\it intrinsic}
variation in the primordial abundances of the light elements,
especially deuterium. Our calculations have predicted that such
PIB-like models could produce light element abundance variations
on baryon mass scales up to ${10}^{10}$ M$_{\odot}$ to
 ${10}^{12}$ M$_{\odot}$.
\vskip 0.1in
\centerline{\bf Acknowledgements}
\vskip 0.1in
We would like to acknowledge useful conversations with Craig Hogan
and Grant
Mathews.
K.J. wishes to acknowledge the hospitality of the Department of Astronomy and
Astrophysics at the University of Chicago.
This work was supported in part by NSF Grant PHY91-21623. It
was
also performed in part under the auspices of the US Department of
Energy
by the Lawrence Livermore National Laboratory
under contract number W-7405-ENG-48.

\vfill\eject
\centerline{\bf 5. References}
\vskip 0.1in

\item{} Alcock, C.R., Dearborn, D.S.P., Fuller, G.M., Mathews, G.J., \&
Meyer, B. 1990, Phys.Rev.Lett., 64, 2607
\item{} Cen, R., Ostriker, P., \& Peebles, P.J.E. 1993, 415, 423
\item{} Chaboyer, B. \& Demarque, P. 1994, ApJ, in press
\item{} Chaboyer, B., Demarque, P., \& Pinsonneault, M.H. 1994, preprint
astro-ph/9408058
\item{} Chaboyer, B., Demarque, P., \& Pinsonneault, M.H. 1994, preprint
astro-ph/9408059
\item{} Chiba, T., Sugiyama, N., \& Suto, Y. 1994, preprint UTAP-93-165
\item{} Davis, M. \& Peebles, P.J.E. 1983, ApJ, 267, 465
\item{} Deliyannis, C.P., Demarque, P., \& Kawaler, S.D. 1990, ApJS, 73, 21
\item{} Dolgov, A. \& Silk, J. 1993, Phys.Rev. D, 47, 4244
\item{} Efstathiou, G., Frenk, C.S., White, S.D.M., \& Davis, M. 1988, MNRAS,
235, 715
\item{} Epstein, R.I. \& Petrosian, V. 1975, ApJ, 197, 281
\item{} Fuller, G.M., Woosley, S.E., \& Weaver, T.A. 1986, ApJ, 307, 675
\item{} Fuller, G.M., Boyd, R., \& Kalen, J. 1991, ApJL, 371, L11
\item{} Gnedin, N.Y. \& Ostriker, J.P. 1992, ApJ, 400, 1
\item{} Gnedin, N.Y., Ostriker, J.P., \& Rees, M.J. 1994, preprint
\item{} Gorski, K.M. \& Silk, J. 1989, ApJL, 346, L1
\item{} Harrison, E.R. 1968, A.J., 73, 533
\item{} Hobbs, L. \& Thorburn, J.A. 1994, ApJL, 428, L25
\item{} Hogan, C.J. 1978, MNRAS, 185, 889
\item{} Hogan, C.J. 1993, ApJL, 415, L63
\item{} Hu, W. \& Sugiyama, N. 1994, preprint CfPA-TH-94-16
\item{} Jedamzik, K. \& Fuller, G.M. 1994, ApJ, 423, 33
\item{} Jedamzik, K., Fuller, G.M., \& Mathews, G.J. 1994, ApJ, 423, 50
\item{} Jedamzik, K., Mathews, G.J., \& Fuller, G.M. 1994, ApJ, in press
\item{} Jedamzik, K. 1994, preprint astro-ph/9408080
\item{} Kashlinsky, A. \& Rees, M.J. 1983, 205, 955
\item{} Kawano, L.H. 1992, preprint Fermilab-Pub-92/04-A
\item{} Kolb, E.W. \& Turner, M.S. 1990, in {\it The Early Universe}
(Reading: Addison-Wesley), pg 321ff
\item{} Kurki-Suonio, H., Matzner, R.A., Centrella, J., Rothman, T., \&
Wilson, J.R. 1988, Phys.Rev. D, 42, 1047
\item{} Kurki-Suonio, H., Matzner, R.A., Olive, K.A., \& Schramm, D.N. 1990,
ApJ, 353, 406
\item{} Kurki-Suonio, H., Jedamzik, K., \& Mathews, G.J. 1994, in preparation
\item{} Loeb, A. 1993, ApJ, 403, 542
\item{} Mathews, G.J., Meyer, B.S., Alcock, C.R., \& Fuller, G.M. 1990, ApJ,
358, 36
\item{} Mathews, G.J., Boyd, R., \& Fuller, G.M. 1993, ApJ, 403, 65
\item{} Meyer, B.S., Alcock, C.R., Mathews, G.J., \& Fuller, G.M. 1990,
Phys.Rev. D, 43, 1079
\item{} Olive, K.A. \& Steigman, G. 1994, preprint UMN-TH-1230/94
\item{} Pagel, B.E.J., Simonson, E.A., Terlevich, R.J., \& Edmunds, M.G.
1992, MNRAS, 255, 325
\item{} Peebles, P.J.E. 1971, {\it Physical Cosmology} (Princeton: Princeton
Univ. Press)
\item{} Peebles, P.J.E. 1987a, ApJL, 315, L73
\item{} Peebles, P.J.E. 1987b, Nature, 327, 210
\item{} Pinsonneault, M.H., Deliyannis, C.P., \& Demarque, P. 1992, ApJS, 78,
179
\item{} Press, W.H. \& Schechter, P. 1974, ApJ, 187, 425
\item{} Rees, M.J. 1984, in {\it Formation and Evolution of Galaxies and
Large Structures in the Universe}, editor D.Reidel Publishing Comp.,pg 271
\item{} Sale, K.E. \& Mathews, G.J. 1986, ApJL, 309, L1
\item{} Sasselov, D. \& Goldwirth, D.S. 1994, ApJ, in press
\item{} Skillman, E.D., Kennicut Jr., R.C. 1993, ApJ, 411, 655
\item{} Skillman, E.D., Terlevich, R., \& Garnett, D.R. 1994, ApJ, in press
\item{} Smith, V.V., Lambert, D.L., \& Nissen, P.E. 1982, ApJ, 408, 262
\item{} Smith, M.S., Kawano, L.H., \& Malaney, R.A. 1993, ApJS, 85, 219
\item{} Songaila, A., Cowie, L.L., Hogan, C.J., \& Rugers, M. 1994, Nature,
368, 599
\item{} Spite, F. \& Spite, M. 1982, A\& A, 115, 357
\item{} Suginohara, T. \& Suto, Y. 1992, ApJ, 387, 431
\item{} Thomas, D., Schramm, D.N., Olive, K.A., Meyer, B., Mathews, G., \&
Fields, B. 1994, ApJ, 430, 291
\item{} Thorburn, J.A. 1994, ApJ, 421, 318
\item{} Wagoner, R.V., Fowler, W.A., \& Hoyle, F. 1967, ApJ, 148, 3
\item{} Walker, T.P., Steigman, G., Schramm, D.N., Olive, K.A., \& Kang,
H.-S. 1991, ApJ, 376, 51
\item{} Yamamoto, K., Nagasawa, M., Sasaki, K., Suzuki, H., Yokoyama, J.
1992, Phys.Rev. D, 46, 4206
\item{} Yokoyama, J. \& Suto, Y. 1991, 379, 427
\item{} Zeldovich, Ya. B. 1975, Soviet Astr. Lett., 1, 5

\vfill\eject

\vfill\eject
\centerline{\bf 6. Figure Captions}
\vskip 0.1in
\item{\bf Figure 1a} A one-dimensional baryon-to-photon number
distribution $\eta
(x)$ divided by the average baryon-to-photon ratio $<\eta >$ as a
function of
space coordinate $x$ (solid line). This distribution has been
generated by
employing the transformation equation (8a) to a gaussian random
variable. The
dotted boxes indicate those regions which are overdense on average by
the
critical amount $\dc $. In this figure we have used $\dc =1.5$. Note
that the
full simulation-\lq\lq volume\rq\rq\ extends from $x=0$ to $x=1$. The
figure
shows only a small part of the full distribution.

\item{\bf Figure 1b} The notation in this figure is as in Figure 1a.
The
distribution shown in this figure has been generated by employing the
transformation equation (8b) to a gaussian random variable. For this
figure we
have used a critical overdensity of $\dc =2$.

\item{\bf Figure 1c} The notation in this figure is as in Figure 1a.
The
distribution shown in this figure has been generated by employing the
transformation equation (8c) to a gaussian random variable. For this
figure
we have used a critical overdensity of $\dc =2$.

\item{\bf Figure 2a} Probability distribution functions
$P_{\lambda}(\eta )$ to
find a region of size $\lambda$ (or, equivalently, mass $M$) with
average
baryon-to-photon ratio $\eta$. These distribution functions are
plotted as a
function of the ratio $\eta /<\eta >$, where $<\eta >$ is the cosmic
average
baryon-to-photon ratio. Distribution functions are shown for the mass
scales
$M=M_c$ (solid line), $M=10M_c$ (dotted line), $M=100M_c$
(short-dashed
line), and $M=1000M_c$ (long-dashed line). Here $M_c$ is the cutoff
mass
scale below which baryon-to-photon fluctuations are assumed to be
suppressed.
For comparison, the dashed-dotted line shows the analytically
determined
distribution function $P(\eta )$ (equation 11). Note that the
ordinate scales
differently for different distribution functions. The
baryon-to-photon
distribution has been generated by employing the transformation
equation (8a)
to a gaussian random variable.

\item{\bf Figure 2b} The notation in this figure is as in Figure 2a.
Distribution
functions are shown for a baryon-to-phton distribution generated by
employing
transformation equation (8b) to a gaussian random variable.

\item{\bf Figure 2c} The notation in this figure is as in Figure 2a.
Distribution
functions are shown for a baryon-to-photon distribution generated by
employing transformation equation (8c) to a gaussian random variable.

\item{\bf Figure 3a} The variance $\sigma_{\eta}(M)$ as a function of
the ratio
of mass to cutoff mass scale $(M/M_c)$. We have generated the
baryon-to-photon distribution by employing transformation equation
(8a) to a
gaussian random variable. The dotted line shows $\sigma_{\eta}(M)$
for a
distribution where a spectral index $n=0$ of the gaussian random
variable has
been used (equation 7). The dashed line shows $\sigma_{\eta}(M)$ for
a model
where $n=-2.4$ has been used. For comparison we also show
$\sigma_{\eta}(M)$
from equation (15) with $n=0$ (lower solid line) and $n=-1.5$ (upper
solid
line).

\item{\bf Figure 3b} The notation in this figure is as in Figure 1a.
For this
figure we use a
distribution in $\eta$ which is characterized by transformation
equation (8b).

\item{\bf Figure 3c} The notation in this figure is as in Figure 3a.
For this figure we use a distribution in $\eta$ which is
characterized by
transformation equation (8c).

\item{\bf Figure 4a} Convergence of numerical results as a function
of the
number of Fourier modes employed in the simulation. Upper panel - the
ratios
$\h2$ (solid line), $^4$He mass fraction $Y_p$ (dotted line), $\li7$
(short-dashed line), and $(\ooo )$ to their convergent values as
noted in the
figures. Lower panel - the variances $\sigma_{\eta}(M_c)$ (solid
line), and
$20\times \sigma_{\eta}(1000\times M_c)$ (dotted line). In these
simulations
we have used the transformation equation (8a),
spectral index $n=0$, and $\dc =1.5$.

\item{\bf Figure 4b} The notation in this figure is as in Figure 4a.
For this
figure we have used a distribution in $\eta$ characterized by
transformation
equation (8b), spectral index $n=0$, and $\dc =2$.

\item{\bf Figure 5a} Light-element nucleosynthesis yields for a model
distribution of $\eta$ which is characterized by the transformation equation
(8a), spectral index $n=0$, $M_c>M_J^b$, and various values of $\oh$ and
$\dc$. The panel in the upper left-hand corner shows the $^4$He mass fraction
$Y_p$ as a function of $\oh$,
whereas the panel in the lower left-hand corner shows the $\li7$
number ratio as a function of $\oh$.
The panels in the upper and lower right-hand corners
show the $\h2$ and $\he3$ number ratios as a function of $\oh$, respectively.
The dotted line is for $\dc =1.25$, the short-dashed line is for $\dc =1.5$,
the long-dashed line is for $\dc =1.75$, while the dashed-dotted line is
for $\dc =2$. The solid lines give the results of standard homogeneous big
bang nucleosynthesis for comparison.

\item{\bf Figure 5b} The values for $\Omega_{diff}h^2$ as a function of $\oh$
for those models for which the nucleosynthesis yields are shown in Figure 5a.

\item{\bf Figure 6a} Nucleosynthesis yields for a model distribution of
$\eta$ which is characterized by the transformation equation (8b), spectral
index $n=0$, $M_c>M_J^b$, and various $\oh$ and $\dc$. The different lines
show results for models with $\dc =1.5$ (dotted line), $\dc =2$
(short-dashed line), $\dc
=2.5$ (long-dashed line), and $\dc =3$ (dashed-dotted line). The notation in
this figure is as in Figure 5a.

\item{\bf Figure 6b} The values for $\Omega_{diff}h^2$ as a function of $\oh$
for those models for which the nucleosynthesis yields are shown in Figure 6a.

\item{\bf Figure 7a} Nucleosynthesis yields for a model distribution of
$\eta$ which is characterized by the transformation equation (8c), spectral
index $n=0$, $M_c>M_J^b$, and various $\oh$ and $\dc$. The different lines
show results for models with $\dc =1.5$ (dotted line), $\dc =2$
(short-dashed line), $\dc
=2.5$ (long-dashed line), and $\dc =3$ (dashed-dotted line). The notation in
this figure is as in Figure 5a.

\item{\bf Figure 7b} The values for $\Omega_{diff}h^2$ as a function of $\oh$
for those models for which the nucleosynthesis yields are shown in Figure 7a.

\item{\bf Figure 8a} Nucleosynthesis yields for a model distribution of
$\eta$ which is characterized by the transformation equation (8a), spectral
index $n=0$, $M_c>M_J^b$, $\dc =1.5$,
and various $\oh$ and collapse efficiency parameters $f$.
The different lines
show results for models with $f=100\%$ (lower dotted line in
the panels for $Y_p$ and $\li7$),
$f=97.5\%$
(short-dashed line), $f=95\%$
(long-dashed line), $f=90\%$ (dashed-dotted line), and $f=80\%$ (upper dotted
line in the panels for $Y_p$ and $\li7$).
Note that the lower dotted lines in the panels for $Y_p$ and $\li7$ correspond
to the upper dotted lines in the panels for $h2$ and $\he3$, and vice versa.
The notation in this figure is as in Figure 5a.

\item{\bf Figure 8b} The values for $\Omega_{diff}h^2$ as a function of $\oh$
for those models for which the nucleosynthesis yields are shown in Figure 8a.
In this figure the results shown by the lower dotted line correspond to
the results shown by the lower dotted lines in the panels for $Y_p$ and
$\li7$ in Figure 8a.

\item{\bf Figure 9a} Nucleosynthesis yields for a model distribution of
$\eta$ which is characterized by the transformation equation (8b), spectral
index $n=0$, $M_c>M_J^b$, $\dc =2$, and various $\oh$ and
and collapse efficiency parameters $f$. The different lines
show results for models with $f=100\%$ (dotted line), $f=99\%$
(short-dashed line), and $f=97.5\%$
(long-dashed line). The notation in
this figure is as in Figure 5a.

\item{\bf Figure 9b} The values for $\Omega_{diff}h^2$ as a function of $\oh$
for those models for which the nucleosynthesis yields are shown in Figure 9a.

\item{\bf Figure 10a} Nucleosynthesis yields for a model distribution of
$\eta$ which is characterized by the transformation equation (8c), spectral
index $n=0$, $M_c>M_J^b$, $\dc =2$, and various $\oh$ and collapse
efficiency parameters $f$. The different lines
show results for models with $f=100\%$ (dotted line), $f=99\%$
(short-dashed line), and $f=97.5\%$
(long-dashed line). The notation in
this figure is as in Figure 5a.

\item{\bf Figure 10b} The values for $\Omega_{diff}h^2$ as a function of $\oh$
for those models for which the nucleosynthesis yields are shown in Figure 10a.

\item{\bf Figure 11a} Nucleosynthesis yields for a model distribution of
$\eta$ which is characterized by the transformation equation (8a), spectral
index $n=0$, $\dc =1.5$, and various $\oh$ and cutoff mass scales $M_c$.
The different lines
show results for models with $M_c>M_J^b$ (dotted line), $M_c=M_J^b/3$
(short-dashed line), $M_c=M_J^b/6$
(long-dashed line), and $M_c=M_J^b/10$ (dashed-dotted line). The notation in
this figure is as in Figure 5a.

\item{\bf Figure 11b} The values for $\Omega_{diff}h^2$ as a function of $\oh$
for those models for which the nucleosynthesis yields are shown in Figure 11a.

\item{\bf Figure 12a} Nucleosynthesis yields for a model distribution of
$\eta$ which is characterized by the transformation equation (8b), spectral
index $n=0$, $\dc =2$, and various $\oh$ and cutoff mass scales $M_c$.
The different lines
show results for models with $M_c>M_J^b$ (dotted line), $M_c=M_J^b/3$
(short-dashed line), $M_c=M_J^b/6$
(long-dashed line), and $M_c=M_J^b/10$ (dashed-dotted line). The notation in
this figure is as in Figure 5a.

\item{\bf Figure 12b} The values for $\Omega_{diff}h^2$ as a function of $\oh$
for those models for which the nucleosynthesis yields are shown in Figure 12a.

\item{\bf Figure 13a} Nucleosynthesis yields for a model distribution of
$\eta$ which is characterized by the transformation equation (8c), spectral
index $n=0$, $\dc =2$, and various $\oh$ and cutoff mass scales $M_c$.
The different lines
show results for models with $M_c>M_J^b$ (dotted line), $M_c=M_J^b/3$
(short-dashed line), $M_c=M_J^b/6$
(long-dashed line), and $M_c=M_J^b/10$ (dashed-dotted line). The notation in
this figure is as in Figure 5a.

\item{\bf Figure 13b} The values for $\Omega_{diff}h^2$ as a function of $\oh$
for those models for which the nucleosynthesis yields are shown in Figure 13a.

\item{\bf Figure 14a} Nucleosynthesis yields for a model distribution of
$\eta$ which is characterized by the transformation equation (8a),
$M_c>M_J^b$, $\dc =1.5$, various $\oh$, and two different effective
spectral indices $n_{eff}$ for the distribution of $\eta$.
The dotted line is for a model with $n_{eff}\approx 0$, whereas the dashed
line is for a model with $n_{eff}\approx -1.5$.
The notation in
this figure is as in Figure 5a.

\item{\bf Figure 14b} The values for $\Omega_{diff}h^2$ as a function of $\oh$
for those models for which the nucleosynthesis yields are shown in Figure 14a.

\item{\bf Figure 15a} Nucleosynthesis yields for a model distribution of
$\eta$ which is characterized by the transformation equation (8b),
$M_c>M_J^b$, $\dc =2$, various $\oh$, and two different
effective spectral indices $n_{eff}$ for the distribution of $\eta$.
The dotted line is for a model with $n_{eff}\approx 0$, whereas the dashed
line is for a model with $n_{eff}\approx -1.5$.
The notation in
this figure is as in Figure 5a.

\item{\bf Figure 15b} The values for $\Omega_{diff}h^2$ as a function of $\oh$
for those models for which the nucleosynthesis yields are shown in Figure 15a.

\item{\bf Figure 16a} Nucleosynthesis yields for a model distribution of
$\eta$ which is characterized by the transformation equation (8c),
$M_c>M_J^b$, $\dc =2$, various $\oh$, two different effective spectral
indices $n_{eff}$ for the distribution of $\eta$. The notation in
this figure is as in Figure 5a.

\item{\bf Figure 16b} The values for $\Omega_{diff}h^2$ as a function of $\oh$
for those models for which the nucleosynthesis yields are shown in Figure 16a.

\item{\bf Figure 17a} The probability distribution functions
$P_{\lambda}\h2$, $P_{\lambda}\he3$, $P_{\lambda}(Y_p)$, and
$P_{\lambda}\li7$ as defined in the text. The panel in the upper left-hand
corner
shows $P_{\lambda}(Y_p)$, while the panel in the lower left-hand corner shows
$P_{\lambda}\li7$. The panels in the upper and lower right-hand corners
show $P_{\lambda}\h2$ and $P_{\lambda }\he3$, respectively. Each panel
shows four probability distribution functions determined on different scales
$\lambda$. The solid line is for $\lambda \leq\lambda_c$,
the dotted line is for
$\lambda =100\lambda_c$, the short-dashed line is for $\lambda
=1000\lambda_c$, whereas the long-dashed line is for $\lambda
=10000\lambda_c$. The distribution functions in this figure have been
computed from a model spatial distribution in $\eta$ characterized by the
transformation equation (8a), spectral index $n=0$, $M_c>M_J^b$, $\dc =1.5$,
and average baryon-to-photon ratio $<\eta >=6\times 10^{-10}$.

\item{\bf Figure 17b} The notation in this figure is as in Figure 17a. The
probability distribution functions shown here have been computed from a model
spatial distribution of $\eta$ which is characterized by transformation
equation (8b), spectral index $n=0$, $M_c>M_J^b$, $\dc =2$, and $<\eta
>=1.2\times 10^{-9}$.

\item{\bf Figure 17c} The notation in this figure is as in Figure 17a. The
probability distribution functions shown here have been computed from a model
spatial distribution of $\eta$ which is characterized by transformation
equation (8c), spectral index $n=0$, $M_c>M_J^b$, $\dc =2$, and $<\eta
>=3\times 10^{-9}$.

\end